\documentclass[draftcls, onecolumn, 12pt]{IEEEtran}

\usepackage{threeparttable,placeins,makecell,stfloats,cite}
\usepackage{threeparttable,arydshln}
\usepackage{amsmath,amssymb}
\usepackage[dvips]{graphicx}
\usepackage[usenames]{color}
\usepackage{amsfonts}
\usepackage{latexsym}
\usepackage{multirow}
\usepackage{caption}
\usepackage{rotating}
\usepackage{float}
\usepackage[format=hang]{subfig}
\usepackage{url}
\usepackage{cite}

\newtheorem{theorem}{\textit{Theorem}}

\newtheorem{remark}{\textit{Remark}}

\begin{document}

\title{Sparse or Dense: A Comparative Study of Code-Domain NOMA Systems}

\author{Zilong~Liu,~\textit{IEEE Senior Member},~Lie-Liang~Yang,~\textit{IEEE Fellow}
\thanks{Zilong Liu is with School of Computer Science and Electrical Engineering, University of Essex, UK (e-mail: zilong.liu@essex.ac.uk). Lie-Liang Yang is with the School of Electronics and Computer Science,
University of Southampton, UK (e-mail: lly@ecs.soton.ac.uk).}
}

\maketitle

\begin{abstract}
This paper is focused on code-domain non-orthogonal multiple access (CD-NOMA), which is an emerging paradigm to support {massive connectivity} for future machine-type wireless networks. We take a comparative approach to study two types of \textit{overloaded} CD-NOMA, i.e., sparse code multiple access (SCMA) and dense code multiple access (DCMA), which are distinctive from each other in terms of their codebooks having \textit{sparsity} or not. By analysing their individual diversity orders (DO) in Rayleigh fading channels,  it is found that DCMA can be designed to enjoy full DO which is equal to the maximum number of resource nodes in the system. This is in contrast to SCMA whose error rate suffers from limited DO equal to the codebook sparsity (i.e., the effective number of resource nodes occupied by each user). We conduct theoretical analysis for the codebook design criteria and propose to use generalized sphere decoder for DCMA detection. We numerically evaluate two types of multiple access schemes under ``$4\times 6$" (i.e., six users communicate over four subcarriers) and ``$5\times 10$" NOMA settings and reveal that DCMA gives rise to significantly improved error rate performance in Rayleigh fading channels, whilst having decoding complexity comparable to that of SCMA.
\end{abstract}


\section{Introduction}\label{section-I}
\subsection{Background}
\IEEEPARstart{T}he trend is that wireless networks have been rapidly evolving towards providing machine-centric data services. Against an increasingly congested and fragmented spectrum, a major research theme nowadays is how to design efficient multiple access protocols to support explosive growth of communication devices. These devices, widely present in a broad range of vertical industries such as factories of future, intelligent refineries and chemical plants, vehicle-to-everything networks, may be densely deployed in certain area for a highly diverse range of data collection and/or control operations. By proper configuration, the devices are mostly in sleep mode with the exception of short periods of time, during which small data packets are exchanged in a sporadic way. The communications over such massive number of communication devices are called machine-type communications (MTC).

An emerging paradigm for MTC networks is called non-orthogonal multiple access (NOMA) which allows \textit{overloaded} multiuser communications, hence enabling a higher spectral efficiency \cite{Dai2015,Liu2017,Cai2018}. Existing NOMA techniques may be mainly categorized into two classes: power-domain NOMA (PD-NOMA) \cite{Saito2013,Higuchi2015,islam2017pdma} and code-domain NOMA (CD-NOMA) \cite{Wang1999,Brannstrom2002,Li2006,Chen2017,Dobre2018}. The former advocates the superposition of two or more users which are assigned with different power levels over the identical time-frequency resources, whereas the latter relies on carefully designed channel codes, interleavers, and codebooks/sequences to separate multiple users. This paper is focused on CD-NOMA systems using different codebooks/sequences.

\subsection{Related Works}
Numerous CD-NOMA schemes have been proposed in recent years. An important research direction of CD-NOMA is to design sequences or codebooks that exhibit certain \textit{sparsity}. In 2008, low-density signatures (LDS) based CDMA was proposed, in which {multiuser detection (MUD) is conducted based on the message passing algorithm (MPA) by efficiently exploiting the sparsity of LDS~\cite{hoshyar2008novel}}. In an LDS-CDMA system, each user spreads its data symbols by a unique LDS whose sequence entries are zero except for a very small fraction. Subsequently, the concept of LDS-CDMA was extended to sparse code multiple access (SCMA), where each user sends a sparse codeword (from a properly designed sparse codebook) according to the instantaneous input message \cite{nikopour2013scma}. Most existing works on SCMA codebook design start from a single multi-dimensional mother constellation having large minimum product distance (or minimum Euclidean distance) \cite{Boutros1996,Boutros1998,Bao2017,Vamegh2019}, with which multiple sparse codebooks are produced through a series of constellation operations, such as phase rotations, interleaving, and permutations \cite{Taherzadeh2014SCMA}. These operations lead to power-imbalanced constellations, i.e., variation of user powers can be seen from sparse codebooks pertinent to each resource node. Power-imbalanced constellations amplify the ``near-far effect" which in turn helps strengthen the interference cancellation/suppression in MPA. The error rate performance of SCMA benefits from the so-called ``constellation shaping gain" (owned by its sparse codebooks).

{It is noted that traditional code-division multiple access (CDMA) \cite{Proakis2008} typically employs non-orthogonal spreading sequences. It belongs to an important class of CD-NOMA~\cite{8352616}, in which the receiver exploits the low cross-correlation properties of spreading sequences for mitigation/suppression of multiuser interference.} Besides SCMA, CDMA has inspired a series of sequence based CD-NOMA proposals in 3GPP discussions \cite{Yang2017}, such as, multiuser shared access (MUSA) \cite{Yuan2016}, non-orthogonal coded access (NOCA) \cite{Nokia2016}, non-orthogonal coded multiple access (NCMA) \cite{LG2016}, and resource spread multiple access (RSMA) \cite{RSMA2016}.  These CD-NOMA schemes may be regarded as dense code multiple access (DCMA) as their sequences (in contrast to that of SCMA) are in general dense, i.e., most\footnote{For example, ternary sequences over $\{0,\pm 1\}$ are adopted in MUSA \cite{Yuan2016}.} or all sequence entries are non-zero.

\subsection{Motivations and Contributions}
Although SCMA has attracted tremendous research attention over the past decade, a comprehensive comparison with DCMA is still lacking, to the best of our knowledge. It is shown in \cite{Lim2017} that the maximum diversity order (DO) of any SCMA system is limited to the sparsity given to each user, i.e., the effective number of resource nodes occupied by each user (denoted by $d_v$). This may fundamentally limit the error rate performance of SCMA systems. By increasing $d_v$ to its maximum, it is intriguing to understand the performance of DCMA (in comparison to SCMA), which has the potential of achieving full DO in Rayleigh fading channels\footnote{After the first review of this work, one reviewer pointed out \cite{Sparse-NOMA-2018} which shows that the spectral efficiency of regular sparse CD-NOMA outperforms DCMA under the setting of AWGN channel and with Gaussian inputs. Unlike \cite{Sparse-NOMA-2018}, however, this paper will mainly investigate and compare the respective DOs as well as the error rate probabilities of DCMA and SCMA in Rayleigh fading channels with finite alphabet inputs. }.

The main contributions of this work are summarized as follows:
\begin{enumerate}
    \item Based on the pairwise error probability (PEP), we analyse the DO of DCMA with an emphasis on $M=4$, where $M$ denotes the number of codewords in each codebook (or the alphabet size in traditional CDMA system), and propose its codebook design criteria in Section~\ref{section-III}. Over uplink Rayleigh fading channels and by applying the inequality of arithmetic and geometric means to the minimum product distance associated to PEP, it is revealed that a) unimodular dense sequences\footnote{A unimodular sequence refers to a sequence whose entries all have identical magnitude. For example, any polyphase sequence is also a unimodular sequence.} lead to optimum DCMA codebooks with full DO, and b) sparse sequences whose nonzero elements are unimodular give rise to an LDS-CDMA enjoying minimum single-error PEPs among all possible SCMA codebooks.

    \item For DCMA transmission over downlink Rayleigh fading channels, we propose to construct dense codebooks based on several celebrated transform matrices in the theory of lattice constellation shaping \cite{Boutros1996,Boutros1998} and multiple-input and multiple-output (MIMO) precoding \cite{Xin2003}. This leads to the DCMA systems which can enjoy the full DO that may not be attainable by random dense codebooks.

    \item To achieve the full error-rate performance of DCMA, we view the system equation as a rank-deficient MIMO system and then carry out non-linear MUD by a generalised sphere decoder (GSD) \cite{Cui2005}. This is different from an iterative LMMSE detector whose system performance heavily relies on the soft message information from a strong channel decoder \cite{Wang1999,Li2006}.

    \item We compare the performances of DCMA and SCMA in terms of their error rates and receiver complexities. Aiming for MTC networks, we evaluate the block error rate (BLER) performances in short packet transmission scenarios. In particular, it is found that a DCMA under GSD enjoys significantly improved error rate performances compared to a corresponding SCMA system with MPA-assisted MUD or a corresponding DCMA system with LMMSE detection\footnote{For CDMA with generalized Welch-bound-equality (WBE) sequences, it is  noteworthy that as shown in \cite{Kapur2005}, the asymptotical error rate of at least one user ``floors" under LMMSE receiver and its full potential can only be attained by nonlinear detection.}.
\end{enumerate}

\subsection{Notations}
$\|\mathbf{X}\|=\sqrt{\sum_{m=1}^{M}\sum_{n=1}^{N}|x_{m,n}|^2}$ stands for the Frobenius norm of matrix $\mathbf{X}=[x_{m,n}]_{m,n=1}^{M,N}$ which is of order $M\times N$. $\text{Tr}(\mathbf{X})$ denotes the trace of square matrix $\mathbf{X}$. $\mathbf{X}^T$ and $\mathbf{X}^H$ denote the transpose and the Hermitian transpose of matrix $\mathbf{X}$. $\text{diag}(\mathbf{x})$ gives a diagonal matrix with the diagonal vector of $\mathbf{x}$. $\mathbf{I}_N$ denotes the identity matrix of order $N$. $|x|$ returns the absolute value of $x$.


\vspace{0.2in}
\section{PRELIMINARIES}\label{section-II}

\subsection{System Model}
 \begin{figure*}[htbp]
  \centering
  \includegraphics[width=6in]{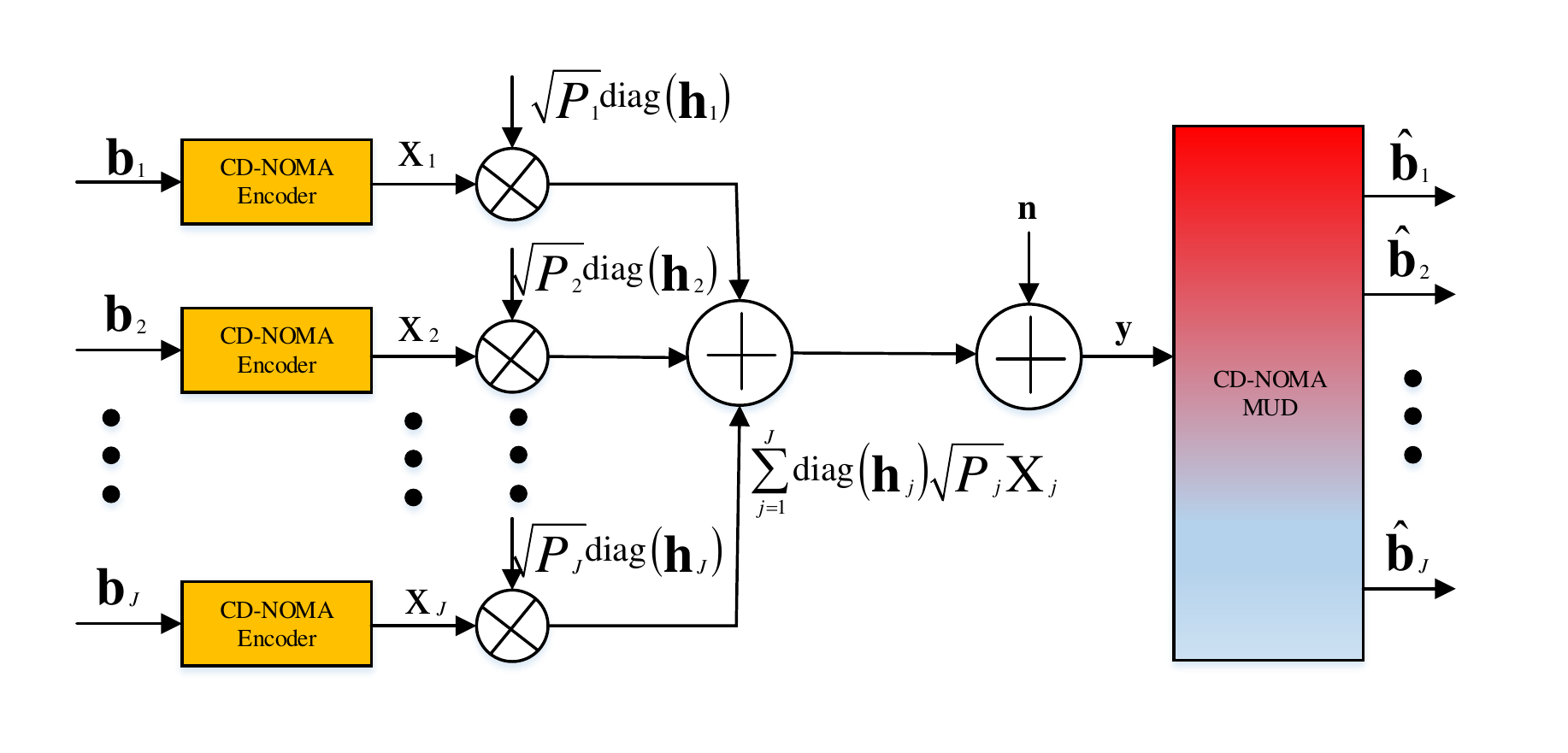}\\
  \caption{A generic CD-NOMA uplink system model with $J$ users of each having different power level $P_j$ ($1\leq j \leq J$).}
  \label{system_model}
\end{figure*}

We consider the uplink $(K\times J)$ CD-NOMA system model as shown in Fig. \ref{system_model}, where $J$ users {(each equipped with single-antenna)} conduct multiple-access communications over $K$ resource nodes. Such a system model can be easily revised to accommodate the downlink channels which we will also study in this work. We adopt orthogonal frequency-division multiplexing (OFDM) to transmit CD-NOMA signals and hence each resource node may also be referred to as a subcarrier channel. By inserting a cyclic prefix before each OFDM symbol, the asynchronous time-offsets in uplink channel can be circumvented. To support massive connectivity in MTC networks, we are particularly interested in designing an overloaded CD-NOMA system with $J>K$, meaning that the number of users that can be simultaneously transmitted is larger than the total number of orthogonal resources. Let $\mathbf{h}_j=[h_{j,1},h_{j,2},\cdots,h_{j,K}]^T$ be the channel fading vector associated to user $j$, where $h_{j,k}\sim\mathcal{CN}(0,1)$. Assume that all the channel fading vectors are uncorrelated and consist of independent complex Gaussian random variables with zero mean and unit variance. Moreover, denote by $\mathbf{n}=[n_1,n_2,\cdots,n_K]^T$ the additive white Gaussian noise (AWGN) vector with $n_k\sim\mathcal{CN}(0,N_0)$. Each user is given a codebook consisting of $M$ codewords with dimension of $K$. Such a codebook may be arranged as a $K\times M$ matrix, denoted by $\mathcal{X}_j,j\in \{1,2,\cdots,J\}$. {Each codebook, sparse or dense, satisfies $\text{Tr}\left ( \mathcal{X}_j \mathcal{X}^{\text{H}}_j\right )=M$.} The CD-NOMA encoder for user $j$ selects a codeword, denoted by $\text{X}_j=[X_{j,1},X_{j,2},\cdots,X_{j,K}]^T$, which is essentially a column of $\mathcal{X}_j$, based on the instantaneous input message $\text{b}_j$ consisting of $\log_2(M)$ bits. {Assume that the total transmit power is $P$ and let $P_j$ ($1\leq j \leq J$) be the transmit power of user $j$ which satisfies $P_j\leq \frac{P}{J}$.} Therefore, the $K$-dimensional received signal $\mathbf{y}$ can be expressed as
\begin{equation}\label{system_equ_uplk}
\mathbf{y}=\sum\limits_{j=1}^J \text{diag}(\mathbf{h}_j)\sqrt{P_j}\text{X}_j+\mathbf{n}.
\end{equation}

In the case of downlink channel, let $\mathbf{n}_j$ be the noise vector seen by user $j$. Thus, the received signal $\mathbf{y}_j$ at user $j$ can be written as
\begin{equation}
\mathbf{y}_j=\text{diag}(\mathbf{h}_j)\cdot \sum\limits_{j=1}^J \sqrt{P_j}\text{X}_j+\mathbf{n}_j,~j=1,2,\cdots,J.
\end{equation}
{For the downlink case, let us assume $\sum_{j=1}^J P_j =P$.}

For every CD-NOMA transmission (donwlink or uplink), all the  codewords from the $J$ users, {upon involving the effect of individual transmit powers},  can be fully expressed as the transmit matrix (TM) below:
\begin{equation}\label{TxMatrix_equ}
    \mathbf{X}=\left[\sqrt{P_1}\text{X}_1,\sqrt{P_2}\text{X}_2,\cdots,\sqrt{P_J}\text{X}_J \right ]_{K\times J}.
\end{equation}
{Note that, we will frequently use TM for DO analysis in Section~\ref{section-III}.}


\subsection{Introduction to SCMA}

 \begin{figure*}[htbp]
  \centering
  \includegraphics[trim=0.2cm 0.2cm 0.2cm 0.2cm, clip=true, width=5in]{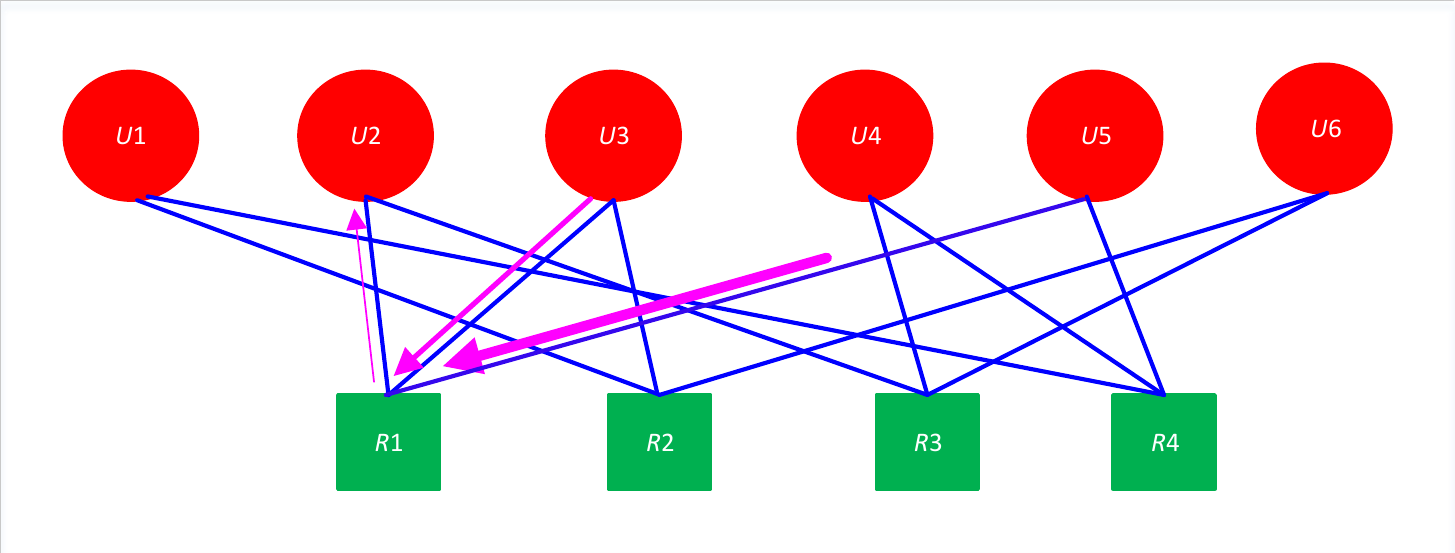}\\
  \caption{Factor graph for an SCMA system with $J=6,K=4,d_v=2,d_c=3$.}
  \label{factor-graph}
\end{figure*}

Sparse codebooks of an SCMA system can be characterized by a bipartite factor graph consisting of resource nodes and user nodes. In this work, we consider the SCMA systems with regular factor graphs, in which each user node has degree of $d_v$ and each resource node has degree of $d_c$. Due to the sparsity, each codeword in $\mathcal{X}_j$ is comprised of $K-d_v$ zeros and $d_v$ non-zero elements .

Fig. \ref{factor-graph} illustrates the factor graph of an SCMA codebook with $J=6,K=4,d_v=2,d_c=3$. In Fig. \ref{factor-graph}, each circle (in green) represents a user node, while each square box (in red) represents a resource node. The arrows (in purple) in Fig. \ref{factor-graph} show the soft messages passed from user nodes to resource node 1 (i.e., $R1$) during MPA decoding at the receiver. An alternative method of representing the factor graph is by an indicator matrix, in which each row is associated to a specific resource node and all the non-zero entries in such a row correspond to the users which have active transmissions over this resource node. Following this principle, the factor graph in Fig. \ref{factor-graph} can be represented by the indicator matrix as follows:
\begin{equation}\label{ind_matrix_equ}
\mathbf{F}=\left [
\begin{matrix}
0 & 1 & 1 & 0 & 1 & 0\\
1 & 0 & 1 & 0 & 0 & 1\\
0 & 1 & 0 & 1 & 0 & 1\\
1 & 0 & 0 & 1 & 1 & 0
\end{matrix}
\right ].
\end{equation}	
In view of the TM defined in (\ref{TxMatrix_equ}), an SCMA system has the following TM structure.
\begin{equation}\label{TM_4x6SCMA}
\mathbf{X}=\left [
\begin{matrix}
0 & \sqrt{P_2}X_{2,1} & \sqrt{P_3}X_{3,1} & 0 & \sqrt{P_5}X_{5,1} & 0\\
\sqrt{P_1}X_{1,2} & 0 & \sqrt{P_3}X_{3,2} & 0 & 0 & \sqrt{P_6}X_{6,2}\\
0 & \sqrt{P_2}X_{2,3} & 0 & \sqrt{P_4}X_{4,3} & 0 & \sqrt{P_6}X_{6,3}\\
\sqrt{P_1}X_{1,4} & 0 & 0 & \sqrt{P_4}X_{4,4} & \sqrt{P_5}X_{5,4} & 0
\end{matrix}
\right ].
\end{equation}

{For given $J$ and $K$, roughly speaking, a larger overloading factor $J/K$ implies a worse error rate probability due to the increase of multiuser interference experienced by each user. By counting the total number of edges in the corresponding factor graph, we have $Jd_v=K d_c$, where $d_v$ and $d_c$ should be carefully chosen in order to maintain the \textit{sparsity} of SCMA system. The sparsity may be ensured if the corresponding factor graph (determined by $J,K,d_v,d_c$) has the minimum cycle\footnote{A cycle in a factor graph is formed by several edges involving user nodes and resource nodes.} size of 6. This can be seen from Fig. \ref{factor-graph} for the $(4\times 6)$-SCMA system (with $d_v=2,d_c=3$). However, when we increase $d_v$ from 2 to a larger value (e.g., 3), the sparsity structure does not hold anymore and the minimum cycle size of the resultant factor graph will be reduced to 4. The latter would result in highly correlated belief messages in MPA decoding and therefore the deterioration of BER.

So far, the design of optimal SCMA codebook remains an open problem. In Subsection III-B, we will show how to design optimal SCMA codebooks in uplink Rayleigh fading channels for $M=4$.
}

\vspace{0.2in}
\section{Analysis and Design of DCMA Systems}\label{section-III}

In this section, we first analyse the DO of DCMA systems based on PEP of every two distinctive TMs with an emphasis on uplink Rayleigh fading channels. For $M=4$, we prove that the optimal codebooks in uplink channels are unimodular sequences. Then, we present codebook selection for {downlink case} as well as DCMA receiver design based on GSD.

\subsection{Analysis of Diversity Order}
\subsubsection{Uplink Channel}
{For a TM $\mathbf{X}$, due to multiuser interference and additive white Gaussian noise, it is assumed to be erroneously decoded to another $K\times J$ matrix $\hat{\mathbf{X}}$, $\hat{\mathbf{X}}\neq\mathbf{X}$, at the receiver,} i.e.,
\begin{equation}
    \hat{\mathbf{X}}=\left[\sqrt{P_1}\hat{\text{X}}_1,\sqrt{P_2}\hat{\text{X}}_2,\cdots,\sqrt{P_J}\hat{\text{X}}_J \right]_{K\times J}.
\end{equation}
{Here, $\hat{\mathbf{X}}$ should be a valid TM, meaning that it is comprised of a combination of multiple valid codewords.} In the sequel, we analyse the PEP conditioned on the channel fading vectors of the uplink channels.
Let us define the element-wise distance $\tau_{j,k}\triangleq \sqrt{P_j}(X_{j,k}-\hat{X}_{j,k})$ and $\hat{\tau}_{j,k}\triangleq (X_{j,k}-\hat{X}_{j,k})$, i.e., we have $\tau_{j,k}=\sqrt{P_j}\hat{\tau}_{j,k}$. Furthermore, let us define
\begin{equation}
    \delta_k \triangleq \sum\limits_{j=1}^{J}h_{j,k}\tau_{j,k},~\Delta_k \triangleq \sqrt{\sum\limits_{j=1}^{J}|\tau_{j,k}|^2}.
\end{equation}
Then, it can be shown that we have
\begin{equation}\label{PEP_uplk}
    \text{Pr}(\mathbf{X}\rightarrow \hat{\mathbf{X}}|\mathbf{h}_j,1\leq j\leq J)=Q\left (\sqrt{\frac{\left \| \sum\limits_{j=1}^J \text{diag}(\mathbf{h}_j)\sqrt{P_j}(\text{X}_j-\hat{\text{X}}_j)\right \|^2}{2N_0}}\right )=
    Q\left (\sqrt{\frac{\sum\limits_{k=1}^{K}|\delta_k|^2}{2N_0}}\right ),
\end{equation}
where $Q(x)=(2\pi)^{-1/2}\int_{x}^{+\infty}e^{-t^2/2}dt$ denotes the tail probability of the standard Gaussian distribution. By \cite{Kim2008}, we have the approximation\footnote{One may also upper bound $Q(x)$ by applying the Chernoff bound, i.e., $Q(x)\leq \exp(-x^2/2)$. But it is relatively loose compared to the approximation of (\ref{Q_equ}). }
\begin{equation}\label{Q_equ}
    Q(x)\simeq \frac{1}{12}\exp(-x^2/2)+\frac{1}{6}\exp(-2x^2/3).
\end{equation}
Applying (\ref{Q_equ}) into (\ref{PEP_uplk}), we obtain
\begin{equation}\label{PEP_uplk2}
    \text{Pr}(\mathbf{X}\rightarrow \hat{\mathbf{X}}|\mathbf{h}_j,1\leq j\leq J)\simeq
    \frac{1}{12}\exp \left (-\frac{\sum\limits_{k=1}^{K}|\delta_k|^2}{4N_0}\right )+\frac{1}{6}\exp \left (-\frac{\sum\limits_{k=1}^{K}|\delta_k|^2}{3N_0}\right ).
\end{equation}
Upon taking the expectation on both sides of (\ref{PEP_uplk2}) and following the derivation similar to that in \cite{Lim2017}, we arrive at
\begin{equation}\label{PEP_equ}
    \text{Pr}(\mathbf{X}\rightarrow \hat{\mathbf{X}})\simeq
    \frac{1}{12}\prod_{k=1}^{K} \frac{1}{1+\frac{\Delta^2_k}{4N_0}}+\frac{1}{6}\prod_{k=1}^{K} \frac{1}{1+\frac{\Delta^2_k}{3N_0}}.
\end{equation}
To proceed, let us define
\begin{equation}
\begin{split}
D(\mathbf{X}\rightarrow \hat{\mathbf{X}}) & \triangleq \left \{k: \Delta_k\neq 0, 1\leq k \leq K \right \},\\
G_d (\mathbf{X}\rightarrow \hat{\mathbf{X}})&\triangleq  \sum\limits_{k=1}^K\text{Ind}\left (\Delta_k\right ),
\end{split}
\end{equation}
where $\text{Ind}(x)$ takes the value of one if $x$ is nonzero and zero otherwise. Clearly, $G_d (\mathbf{X}\rightarrow \hat{\mathbf{X}})$ gives the cardinality of set $D(\mathbf{X}\rightarrow \hat{\mathbf{X}})$.
In high SNR region, we have $1+{\Delta^2_k}/{4N_0}\approx {\Delta^2_k}/{4N_0}$ and $1+{\Delta^2_k}/{3N_0}\approx {\Delta^2_k}/{3N_0}$. Thus, the PEP in (\ref{PEP_equ}) may be written as 
\begin{equation}\label{PEP_equ2}
 \text{Pr}(\mathbf{X}\rightarrow \hat{\mathbf{X}})\simeq (1/N_0)^{-G_d (\mathbf{X}\rightarrow \hat{\mathbf{X}})} \cdot \left ( \frac{4^{-G_d (\mathbf{X}\rightarrow \hat{\mathbf{X}})}}{12}+\frac{3^{-G_d (\mathbf{X}\rightarrow \hat{\mathbf{X}})}}{6}\right ) \cdot \prod_{k\in D (\mathbf{X}\rightarrow \hat{\mathbf{X}})}\Delta_k^{-2} .
\end{equation}
{From \eqref{PEP_equ2}, it is implied that $G_d\triangleq \min_{\mathbf{X}\neq \hat{\mathbf{X}}}G_d (\mathbf{X}\rightarrow \hat{\mathbf{X}})$ is the DO\footnote{It is noted that DO is an important concept in communication theory. For example, it has been widely used in the study of space-time coding where DO arises due to the use of multiple antennas. Formally, DO is defined as $\text{DO}=\lim_{\text{SNR}\rightarrow \infty}-\frac{\log \text{BER}}{\log \text{SNR}}$. That is, DO measures the number of independent paths/channels over which the data is received. In the context of CD-NOMA system, DO refers to the number of orthogonal resources which allows the transmission of any two distinctive TMs.} of CD-NOMA system.} Based on (\ref{PEP_equ2}), we have the following observations.

\textit{Observation 1}:  {As indicated in (\ref{PEP_equ2}), to achieve the best possible error rate performance, it is desirable to attain the full DO of $G_d=K$. Explicitly, the full DO can be achieved if and only if the codebooks are dense. Correspondingly, the resultant CD-NOMA systems are referred to as DCMA systems. {It should be noted that this does not mean any DCMA system can achieve full DO, unless the corresponding dense codebooks satisfy certain structural properties. Several dense codebooks achieving full DO will be introduced in the end of Subsection III-B.}. }

\textit{Observation 2}: The full DO may never be attained by SCMA systems. In general, the DO of an SCMA system is limited to $d_v$ (i.e., the effective number of resource nodes utilized by each user) which is {usually very small} compared to $K$.  To reveal this, let us take the $4\times 6$ SCMA system shown in Fig. \ref{factor-graph}  as an example. Let us consider $\mathbf{X}$ in (\ref{TM_4x6SCMA}) as the TM, which however may be decoded to the following matrix (with ${X}_{6,2}\neq \hat{X}_{6,2}, {X}_{6,3}\neq \hat{X}_{6,3}$) by the receiver:
\begin{equation}\label{TM_4x6SCMA_example}
\hat{\mathbf{X}}=\left [
\begin{matrix}
0 & \sqrt{P_2}X_{2,1} & \sqrt{P_3}X_{3,1} & 0 & \sqrt{P_5}X_{5,1} & 0\\
\sqrt{P_1}X_{1,2} & 0 & \sqrt{P_3}X_{3,2} & 0 & 0 & \sqrt{P_6}\hat{X}_{6,2}\\
0 & \sqrt{P_2}X_{2,3} & 0 & \sqrt{P_4}X_{4,3} & 0 & \sqrt{P_6}\hat{X}_{6,3}\\
\sqrt{P_1}X_{1,4} & 0 & 0 & \sqrt{P_4}X_{4,4} & \sqrt{P_5}X_{5,4} & 0
\end{matrix}
\right ].
\end{equation}
One can see that the two matrices only differ in the last column, meaning that the decoding errors occurred with the sixth user. In this case, it is easy to show that $G_d (\mathbf{X}\rightarrow \hat{\mathbf{X}})=2$ and consequently the DO of such an SCMA system is $G_d=2$.

\textit{Observation 3}: { From (\ref{PEP_equ2}) we can know that for both SCMA and DCMA systems, an important codebook design criteria is maximizing the product-distance  $\prod_{k\in D (\mathbf{X}\rightarrow \hat{\mathbf{X}})}\Delta_k$ for every pair of $(\mathbf{X},\hat{\mathbf{X}})$, $\mathbf{X}\neq \hat{\mathbf{X}}$.} So far, the optimal codebook design for SCMA systems remains open.

\subsubsection{Downlink Channel}
{Following a similar derivation to the above for the uplink channels, we can obtain the PEP $\text{Pr}^{(j)}(\mathbf{X}\rightarrow \hat{\mathbf{X}})$ ($1\leq j \leq J$), where superscript ``$(j)$" is used to indicate that the PEP analysis is carried out at user $j$. It can be shown that this PEP has the same form as (\ref{PEP_equ2}), but with the definitions of $\delta_{j,k}$ and $\Delta_{j,k}$ respectively given by
\begin{equation}\label{delta_equ2}
    \delta_{j,k} \triangleq  h_{j,k}\cdot\sum\limits_{j=1}^{J}\tau_{j,k},~\Delta_{j,k} \triangleq {\left |\sum\limits_{j=1}^{J}\tau_{j,k} \right |}.
\end{equation}
Furthermore, we can show that the three observations stated above for the uplink channels are also valid for the downlink scenario upon taking $\Delta_{j,k}$ defined in (\ref{delta_equ2}) into account.}

\vspace{0.1in}
\subsection{Design of DCMA System}

Let us consider the linear encoding for a DCMA codebook in the following way:
\begin{equation}
\text{X}_j=\mathbf{G}_j\mathbf{u}_j,~ 1\leq j \leq J,
\end{equation}
where $\mathbf{G}_j=[\mathbf{g}_{j,1},\mathbf{g}_{j,2},\cdots,\mathbf{g}_{j,\log_2M}]$ denotes the generator matrix of user $j$, {which is comprised of $\log_2 M$ complex-valued column vectors $\mathbf{g}_{j,m}(1\leq m \leq \log_2 M)$ of length $K$}, and $\mathbf{u}_j=[u_{j,1},u_{j,2},\cdots,u_{j,\log_2 M}]^T\in \{1,-1\}^{\log_2 M}$ stands for user $j$'s instantaneous input binary message vector. By including all the $\mathbf{u}_j$'s according to their corresponding integer values in ascending order, we form a $\log_2 M \times M$ matrix $\mathbf{U}$. {Let ``$+$" and ``$-$" be $+1$ and $-1$, respectively.} For example, when $M=4$, we have
\begin{equation}
\mathbf{U}=\left [
\begin{matrix}
-+-+\\
--++
\end{matrix}
\right],
\end{equation}
and when $M=16$, we have
\begin{equation}
\mathbf{U}=\left [
\begin{matrix}
-+-+-+-+-+-+-+-+\\
--++--++--++--++\\
----++++----++++\\
--------++++++++
\end{matrix}
\right]_{4\times 16}.
\end{equation}
Therefore, the codebook for user $j$ is $\mathcal{X}_j=\mathbf{G}_j \mathbf{U}$. Based on our earlier assumption that $\text{Tr}\left ( \mathcal{X}_j \mathcal{X}^{\text{H}}_j\right )=M$ (see Subsection II-A), we obtain $\text{Tr}\left ( \mathbf{G}_j \mathbf{G}^{\text{H}}_j\right )=1$, implying that $\sum_{t=1}^{\log_2 M} \|\mathbf{g}_{j,t}\|^2=1$. Assuming that equal error protection is provided to the $\log_2 M$ bits of each codeword, it is natural to have $\|\mathbf{g}_{j,t}\|^2=1/\log_2 M$ for all $1\leq t \leq \log_2 M$.

Define {$n_{\text{E}}(\mathbf{X},\hat{\mathbf{X}})$} as the number of erroneous bits when $\hat{\mathbf{X}}$ is decoded at the receiver. By the union bound, the average bit error rate (ABER) of a CD-NOMA system satisfies
\begin{equation}\label{ABER_equ}
\begin{split}
P_{\text{b}} & \leq \frac{1}{M^J \cdot J\log_2(M)} \cdot\sum_{\mathbf{X}}\sum_{\hat{\mathbf{X}}\neq \mathbf{X}}{n_{\text{E}}(\mathbf{X},\hat{\mathbf{X}})}\cdot\text{Pr}(\mathbf{X}\rightarrow \hat{\mathbf{X}}).
\end{split}
\end{equation}

\subsubsection{Uplink Channel}
We note that the TM error pattern example in \textit{Observation 2} [see (\ref{TM_4x6SCMA_example})] can be categorized into the case that the decoding errors occur with single user only. Such kind of error pattern is called the ``single-error pattern'' in this paper; otherwise, it will be called the ``multiple-error pattern''. Let us write the PEP for a single-error pattern as $\text{Pr}_s(\mathbf{X}\rightarrow \hat{\mathbf{X}})$ and denote by $j_s$ the corresponding error user index.  In literature, it is widely observed that the average error rate performance of any precoded system is dominated by the single-error patterns \cite{Lim2017,Cai2004} in high SNR region. Let $C\triangleq(1/N_0)^{-K} \cdot \left ( \frac{4^{-K}}{12}+\frac{3^{-K}}{6}\right )$. 
In this case, we have $\Delta^2_k = |\tau_{j_s,k}|^2=P_{j_s}|\hat{\tau}_{j_s,k}|^2$, where $1\leq k \leq K$, and
\begin{equation}
\text{Pr}_s(\mathbf{X}\rightarrow \hat{\mathbf{X}})=C\cdot P^{-K}_{j_s} \cdot \prod_{k=1}^K |\hat{\tau}_{j_s,k}|^{-2}.
\end{equation}

For $M=4$, let the generator matrix for user $j_s$ be written as $\mathbf{G}_{j_s}=[\mathbf{g}_{j_s,1},\mathbf{g}_{j_s,2}]$, the transmit codeword of user $j_s$ be $\text{X}_{j_s}=\mathbf{g}_{j_s,1}u_{j_s,1}+\mathbf{g}_{j_s,2}u_{j_s,2}$, whereas the codeword detected by the receiver be $\hat{\text{X}}_{j_s}=\mathbf{g}_{j_s,1}\hat{u}_{j_s,1}+\mathbf{g}_{j_s,2}\hat{u}_{j_s,2}$. Let us define $e_{j_s,1}\triangleq u_{j_s,1}-\hat{u}_{j_s,1}$ and $e_{j_s,2}\triangleq u_{j_s,2}-\hat{u}_{j_s,2}$, where {$e_{j_s,1},e_{j_s,2}\in \{-2,0,2\}$} and at least one of them is nonzero. Then, for the product-distance introduced in \textit{Observation 3}, with the aid of the inequality of arithmetic and geometric means, we have 
\begin{equation}\label{prod_dist_upbd}
\prod_{k\in D (\mathbf{X}\rightarrow \hat{\mathbf{X}})}\Delta^2_k=\prod_{k=1}^{K}|\tau_{j_s,k}|^2\leq \left ( \frac{\sum_{k=1}^K |\tau_{j_s,k}|^2}{K}\right )^K=\frac{P_{j_s}^K}{K^K}\cdot   \left \|\text{X}_{j_s}-\hat{\text{X}}_{j_s } \right\|^{2K},
\end{equation}
where the equality is achieved if and only if $|\hat{\tau}_{j_s,1}|^2=|\hat{\tau}_{j_s,2}|^2=\cdots =|\hat{\tau}_{j_s,K}|^2$. Observe that $\text{X}_{j_s}-\hat{\text{X}}_{j_s }=\mathbf{g}_{j_s,1}e_{j_s,1}+\mathbf{g}_{j_s,2}e_{j_s,2}$. Then, in order to meet the product-distance upper bound in (\ref{prod_dist_upbd}) with equality, we proceed with the discussion according to the following three cases:
\begin{enumerate}
\item If $e_{j_s,1}\neq 0$ and $e_{j_s,2}=0$, we have $\text{X}_{j_s}-\hat{\text{X}}_{j_s }=\mathbf{g}_{j_s,1}e_{j_s,1}$. Hence, it is required that all the elements of $\mathbf{g}_{j_s,1}$ take identical magnitude.
\item If $e_{j_s,1}= 0$ and $e_{j_s,2}\neq 0$, we can show that all the elements of $\mathbf{g}_{j_s,2}$ should take identical magnitude.  One may also apprehend our previous assumption that $\|\mathbf{g}_{j_s,1}\|^2=\|\mathbf{g}_{j_s,2}\|^2$ in order to provide equal error protection to both bits in such a codeword.
\item If both $e_{j_s,1}\neq 0$ and $e_{j_s,2}\neq 0$, by the triangle inequality, we have
\begin{equation}
\left \|\text{X}_{j_s}-\hat{\text{X}}_{j_s } \right\|^2=\left \|\mathbf{g}_{j_s,1}e_{j_s,1}+\mathbf{g}_{j_s,2}e_{j_s,2} \right\|^2\leq \left \|\mathbf{g}_{j_s,1}\right\|^2 + \left \|\mathbf{g}_{j_s,2} \right\|^2=1,
\end{equation}
where the equality is achieved for all the four combinations of $(e_{j_s,1},e_{j_s,2})$, if and only if $\mathbf{g}_{j_s,1}$ and $\mathbf{g}_{j_s,2}$ are perpendicular in multidimensional space, i.e.,  $\mathbf{g}_{j_1,2}=\pm i\mathbf{g}_{j_s,1}$.
\end{enumerate}

\vspace{0.1in}

\begin{remark}\label{Rmk4SinglErrPEP}
When $M=4$ {and for given $P_{j_s}$}, all the single-error PEPs are minimized if and only if unimodular sequence spreading is adopted in a DCMA system, i.e.,  each transmit codeword takes one of the following forms: $\text{X}_{j}=\mathbf{g}_{j,1}(b_{j,1}+ib_{j,2})$ or $\text{X}_{j}=\mathbf{g}_{j,1}(b_{j,1}-ib_{j,2})$, where $b_{j,1},b_{j,2}\in \{-1,1\}$ and $\mathbf{g}_{j,1}$ is a unimodular sequence. In this case, the product distance corresponding to each single-error PEP is maximized.
\end{remark}

\vspace{0.1in}

{For multiple-error patterns, let us assume that there are $m$ users in error and these users' indices are $j_{s_1},j_{s_2},\cdots,j_{s_{m}}$, where $2\leq m\leq J$. Then, with the aid of the inequality of arithmetic and geometric means, and applying the similar analysis as in the case of single-error patterns}, we have
\begin{equation}\label{prod_dist_upbd2}
\prod_{k\in D (\mathbf{X}\rightarrow \hat{\mathbf{X}})}\Delta^2_k\leq \frac{1}{K^K}\cdot   \left (\sum\limits_{t=1}^m{P_{j_{s_t}}}\left \|\text{X}_{j_{s_t}}-\hat{\text{X}}_{j_{s_t} } \right\|^{2} \right )^K,
\end{equation}
where the equality is achieved if and only if
\begin{equation}\label{multipleerr_equality_equ}
\sum\limits_{t=1}^m {P_{j_{s_t}}} |\hat{\tau}_{j_{s_t},1}|^2=\sum\limits_{t=1}^m {P_{j_{s_t}}} |\hat{\tau}_{j_{s_t},2}|^2=\cdots=\sum\limits_{t=1}^m {P_{j_{s_t}}} |\hat{\tau}_{j_{s_t},K}|^2.
\end{equation}
Upon taking into account of the observation made in \textit{Remark \ref{Rmk4SinglErrPEP}}, one can see that (\ref{multipleerr_equality_equ}) is also held when unimodular spreading sequences are adopted for $M=4$.

\vspace{0.1in}
{\noindent{\textit{Power Allocation in Uplink Channel}}: To design an enhanced full-diversity DCMA with $G_d(\mathbf{X}\rightarrow \hat{\mathbf{X}})=K$, it is enlightening to discuss the power allocation in order to minimize the upper bound of (\ref{ABER_equ}). When the SNR is sufficiently high, we can simply consider to minimize the sum of the single-error PEP terms in (\ref{ABER_equ}), all of which have an identical number of erroneous bits ${n_{\text{E}}(\mathbf{X},\hat{\mathbf{X}})}=1$\footnote{Note that, such analysis can be carried out similarly for other value of ${n_{\text{E}}(\mathbf{X},\hat{\mathbf{X}})}$, but this will not change the power allocation scheme present in the sequel.}, i.e.,
\begin{align}\label{SingleUPEP_sum}
\sum_{\mathbf{X}}\sum_{\hat{\mathbf{X}}\neq \mathbf{X}}\text{Pr}_s(\mathbf{X}\rightarrow \hat{\mathbf{X}})
= & \sum\limits_{j_s=1}^J\sum_{\text{X}_{j_s}}\sum_{\text{X}_{j_s}\neq \hat{\text{X}}_{j_s}}\text{Pr}(\text{X}_{j_s}\rightarrow \hat{\text{X}}_{j_s})\nonumber\\
= & C\cdot \sum\limits_{j_s=1}^J P^{-K}_{j_s}\sum_{\text{X}_{j_s}}\sum_{\text{X}_{j_s}\neq \hat{\text{X}}_{j_s}}\prod_{k=1}^K |\hat{\tau}_{j_s,k}|^{-2}.
\end{align}
When unimodular sequences are adopted for $M=4$, all the entries in
\begin{displaymath}
\Bigl \{|\hat{\tau}_{j,k}|:1\leq k \leq K,1\leq j\leq J \Bigl \}
\end{displaymath}
 take an identical value due to the spreading nature of DCMA. Hence, we have
\begin{equation}\label{SingleUPEP_sum2}
\prod_{k=1}^K |\hat{\tau}_{1,k}|^{-2}=\prod_{k=1}^K |\hat{\tau}_{2,k}|^{-2}=\cdots =\prod_{k=1}^K |\hat{\tau}_{J,k}|^{-2},
\end{equation}
and
\begin{equation}\label{SingleUPEP_sum3}
\sum_{\text{X}_{1}}\sum_{\text{X}_{1}\neq \hat{\text{X}}_{1}}\prod_{k=1}^K |\hat{\tau}_{1,k}|^{-2}=\sum_{\text{X}_{2}}\sum_{\text{X}_{2}\neq \hat{\text{X}}_{2}}\prod_{k=1}^K |\hat{\tau}_{2,k}|^{-2}=\cdots =\sum_{\text{X}_{J}}\sum_{\text{X}_{J}\neq \hat{\text{X}}_{J}}\prod_{k=1}^K |\hat{\tau}_{J,k}|^{-2}.
\end{equation}
 Recall that $P_j\leq P/J$ ($1\leq j \leq J$) should be satisfied for uplink channels. Therefore, it can be readily shown that the sum of the single-errror PEP terms in (\ref{SingleUPEP_sum}) is minimized in high SNR region, if and only if $P_1=P_2=\cdots= P_J=P/J$, i.e., if uniform power allocation is employed. Inspired by this observation,
we assume uniform power allocation for all the uplink CD-NOMA systems in the forthcoming discourses.}

\vspace{0.1in}
Based on the above analysis, we introduce the following theorem:

\begin{theorem}\label{main-theorem}
For an uplink quaternary (i.e., $M=4$) DCMA with uniform power allocation, all the PEPs (and hence the ABER) are minimized provided that the dense codebooks are formed by unimodular spreading sequences. In this case, all the product-distance terms in the left-hand-side of (\ref{prod_dist_upbd2}) are maximized.
\end{theorem}

\vspace{0.1in}

\begin{remark}
{When $M=2^q>4$, employing unimodular sequence spreading and $2^q$-QAM modulation ensures a DCMA with full DO, but does not necessarily yield the largest product distance and minimum PEPs.}
\end{remark}

\vspace{0.1in}

\begin{remark}\label{rmk4SCMA}
The assertions in \textit{Remark \ref{Rmk4SinglErrPEP}} also apply to SCMA systems: When $M=4$, all the single-error PEPs are minimized by LDS-CDMA, if and only if all the nonzero elements of sparse sequences take identical magnitude. Due to the sparsity of LDS, however, the same may not be held when multiple-error PEPs are considered.
\end{remark}
\vspace{0.1in}

Denote by $\mathbf{s}_j=[s_{j,1},s_{j,2},\cdots,s_{j,K}]^T$, where $|s_{j,1}|=|s_{j,2}|=\cdots =|s_{j,K}|=1/\sqrt{K}$ for all $1\leq k \leq K$, the dense sequence assigned to user $j$ and $\mathcal{A}=\{\alpha_1,\alpha_2,\cdots,\alpha_M\}$ the alphabet set (with zero mean and unit variance) shared by all the $J$ users.  Hence, the codebook for user $j$ is
\begin{equation}
\mathcal{X}_j=[\alpha_1 \mathbf{s}_j,\alpha_2 \mathbf{s}_j,\cdots, \alpha_M \mathbf{s}_j]_{K\times M}.
\end{equation}
By the above settings, clearly $\text{Tr}\left ( \mathcal{X}_j \mathcal{X}^{\text{H}}_j\right )=M$ holds. Based on \textit{Remarks 1} and \textit{3}, we apply unimodular spreading sequences $\{\mathbf{s}_j:1\leq j\leq J\}$ and uniform power allocation to an uplink DCMA system. Again, we consider $M=4$ and let $\mu_j=(b_{j,1}+ib_{j,2})/\sqrt{2}\in \mathcal{A}=\{\pm 1\pm i\}/\sqrt{2}$ be the current transmit symbol of user $j$, where $\text{b}_j=[b_{j,1},b_{j,2}]^T\in \{-1,1\}^2$.  Let $\text{X}_j=\mathbf{s}_j \mu_j$ and plug it into (\ref{system_equ_uplk}), we obtain
\begin{equation}\label{system_equ_uplk2}
\mathbf{y}=\mathbf{H} \mathbf{u}+\mathbf{n},
\end{equation}
where
\begin{equation}
\begin{split}
\mathbf{H}& =\sqrt{{P}/{J}} \cdot\Bigl [ \text{diag}(\mathbf{h}_1)\mathbf{s}_1,\text{diag}(\mathbf{h}_2)\mathbf{s}_2,\cdots, \text{diag}(\mathbf{h}_J)\mathbf{s}_J\Bigl ]_{K\times J},\\
\mathbf{u}&=[\mu_{1},\mu_{2},\cdots,\mu_{J}]^T\in \mathcal{A}^J.
\end{split}
\end{equation}

{For higher-order modulation, it is noted that one may also obtain a system equation similar to (\ref{system_equ_uplk2}). As an example, let us consider 16-QAM (i.e., $M=16$) and $\mu_j=\frac{2}{\sqrt{5}}\mu^1_j+\frac{1}{\sqrt{5}}\mu^2_j$, where
\begin{displaymath}
\begin{split}
\mu^1_j&=(b^1_{j,1}+ib^1_{j,2})/\sqrt{2},~b^1_{j,1},b^1_{j,2}\in \{-1,1\},\\
\mu^2_j&=(b^2_{j,1}+ib^2_{j,2})/\sqrt{2},~b^2_{j,1},b^2_{j,2}\in \{-1,1\}.
\end{split}
\end{displaymath}
In this case, the updated $\mathbf{H}$ and $\mathbf{u}$ in $\mathbf{y}=\mathbf{H} \mathbf{u}+\mathbf{n}$ can be expressed as follows:
\begin{equation}
\begin{split}
\mathbf{H}& =\sqrt{\frac{P}{5J}} \cdot\Bigl [ 2\text{diag}(\mathbf{h}_1)\mathbf{s}_1,\cdots, 2\text{diag}(\mathbf{h}_J)\mathbf{s}_J,  \text{diag}(\mathbf{h}_1)\mathbf{s}_1,\cdots, \text{diag}(\mathbf{h}_J)\mathbf{s}_J \Bigl ]_{K\times 2J},\\
\mathbf{u}&=[\mu^1_{1},\cdots,\mu^1_{J},\mu^2_{1},\cdots,\mu^2_{J}]^T\in \mathcal{A}^{2J}.
\end{split}
\end{equation}
}

\vspace{0.1in}
\subsubsection{Downlink Channel}
In the above uplink case, random Rayleigh fading coefficients provide a unique way to harvest the full DO of DCMA. However, this may not hold true for the downlink case as all the users superimposed over any resource node experience an identical Rayleigh fading gain. {In this case, proper selection of dense codebooks is required in order to yield large product distances.} Specifically, in order to attain the full DO at user $j$ ($1\leq j \leq J$),  we require that
\begin{equation}\label{Delta_dnlk_equ}
\Delta_{j,k} = {\left |\sum_{j=1}^{J}\tau_{j,k} \right |}>0,~\text{for all}~1\leq k \leq K.
\end{equation}
Rewrite (\ref{system_equ_uplk2}) for the downlink case as:
\begin{equation}\label{system_equ_dnlk}
\mathbf{y}_j=\sqrt{{P}/{J}}\cdot\text{diag}(\mathbf{h}_j)  \mathbf{S} \mathbf{u}+\mathbf{n}_j,
\end{equation}
where
\begin{equation}\label{Su_equ}
\begin{split}
\mathbf{S}& =  \left [ \mathbf{s}_1,\mathbf{s}_2,\cdots, \mathbf{s}_J \right ]=\left[\mathbf{r}^T_1,\mathbf{r}^T_2,\cdots,\mathbf{r}^T_K \right ]^T.
\end{split}
\end{equation}
In (\ref{Su_equ}), $\mathbf{r}_k$ ($1\leq k \leq K$) denotes the $k$-th row of the matrix $\mathbf{S}$ formed by $J$ number of length-$K$ dense sequences. It can be readily shown that (\ref{Delta_dnlk_equ}) is equivalent to
\begin{equation}\label{Delta_dnlk_equ2}
\prod_{k=1}^K \left | \mathbf{r}_k(\mathbf{u}-\hat{\mathbf{u}})\right |>0,~\text{for any}~\hat{\mathbf{u}}\neq {\mathbf{u}}~\text{and}~\hat{\mathbf{u}},{\mathbf{u}}\in \mathcal{A}.
\end{equation}

The spreading matrices $\mathbf{S}$'s satisfying (\ref{Delta_dnlk_equ2}) can be obtained by some good transform matrices provided in the literature on lattice constellation shaping and MIMO linear precoding. Below we summarize some of the best known transform matrices. To this end, let us define the transpose of a $J \times J$ Vandermonde matrix based on variables $\theta_1,\theta_2,\cdots,\theta_J$ as:
\begin{equation}
\Theta(\theta_1,\theta_2,\cdots,\theta_J)\triangleq\frac{1}{\gamma} \left [
\begin{matrix}
1 & \theta_1 & \theta^2_1 & \cdots & \theta^{J-1}_1\\
1 & \theta_2 & \theta^2_2 & \cdots & \theta^{J-1}_2\\
\vdots & \vdots & \vdots & \ddots & \vdots \\
1 & \theta_J & \theta^2_J & \cdots & \theta^{J-1}_J\\
\end{matrix}
\right ],
\end{equation}
where $\gamma$ is the normalization factor to ensure that $\text{Tr}(\Theta \Theta^H)=J^2/K$. The following constructions give {the corresponding} $\theta_j$ for $1\leq j \leq J$.

\noindent \textit{Construction 1}: If $J=2^s$ ($s\geq 1$), we have $\theta_j=\exp\left (i\frac{4j-3}{2J}\pi\right )$ \cite{Boutros1998}.

\noindent \textit{Construction 2}: If $J=2^s\cdot 3^t$ ($s,t\geq 1$), we have $\theta_j=\exp\left (i\frac{6j-5}{3J}\pi\right )$ \cite{Boutros1998}.

\noindent \textit{Construction 3}: If $J\neq 2^s (s\geq 1)$ but $J=\phi (L)$  for $L \neq 0~ (\text{mod}~4)$, we have $\theta_j=\exp\left ( i\frac{2m_j}{L}\pi\right )$, where $\text{gcd}(m_j,L)\footnote{$\text{gcd}(x,y)$ refers to the greatest common divisor (gcd) of integers $x$  and $y$. For example, \text{gcd}(8,12)=4 and \text{gcd}(15,20)=5.}=1,1\leq m_j\leq L$ and $\phi(\cdot)$ denotes the Euler function\footnote{For example, $\phi(7)=6$ as 1, 2, 3, 4, 5, 6 are co-prime with 7. } which refers to the total number of positive integers that are less than and co-prime to the integer argument \cite{Xin2003}.

\noindent \textit{Construction 4}: If $J$ is odd, we have $\theta_j=2^{\frac{1}{2J}}\exp\left (i\frac{8j-7}{4J}\pi\right )$ \cite{Xin2003}.

Denote by $\Theta_k$ the $k$-th row of $\Theta$. Any transform matrix $\Theta$ from the above constructions has the property that $\prod_{k=1}^J \left | \Theta_k(\mathbf{u}-\hat{\mathbf{u}})\right |>0,~\text{for any}~\hat{\mathbf{u}}\neq {\mathbf{u}}~\text{and}~\hat{\mathbf{u}},{\mathbf{u}}\in \mathcal{A}$, where the alphabet set $\mathcal{A}$ is carved from $Z[i]\triangleq \{a+i b: ~a~\text{and}~b~\text{are integers}\}$. In particular, the transform matrices from \textit{Constructions 1} and \textit{3} are optimal in terms of the maximum coding gain \cite{Xin2003}. By randomly selecting $K$ rows out of $\Theta$,  a spreading matrix $\mathbf{S}$ satisfying (\ref{Delta_dnlk_equ2}) is obtained.

\vspace{0.1in}
{\noindent{\textit{Power Allocation in Downlink Channel}:} It is noted that the assertion of \textit{Remark \ref{Rmk4SinglErrPEP}} also holds for the downlink case: when $M=4$ {and for given $P_{j_s}$}, all the single-error PEPs are minimized if and only if unimodular sequence spreading is adopted in a DCMA system. Moreover, it can be verified that (\ref{SingleUPEP_sum})-(\ref{SingleUPEP_sum3}) are also valid. Recall the downlink power constraint $\sum_{j=1}^J P_j=P$ and let $\mathbf{P}=[P_1,P_2,\cdots,P_J]^T$. In order to minimize the sum of the single-error PEP terms in (\ref{ABER_equ}), it is equivalent to minimize the following \textit{Lagrangian dual function}:
\begin{equation}
\begin{split}
f(\mathbf{P},\nu)&=\sum\limits_{j=1}^J P^{-K}_{j} + \nu\left (\sum\limits_{j=1}^J P_j-P\right ).
\end{split}
\end{equation}
By taking the derivative of $f(\mathbf{P},\nu)$ with respect to $P_j(1\leq j \leq J)$ for the optimality condition, we can show that the sum of the single-user PEP terms in (\ref{SingleUPEP_sum}) is minimized in high SNR region, if and only if $P_1=P_2=\cdots= P_J=P/J$, i.e., if uniform power allocation is employed. Therefore, in Section IV, we use uniform power allocation for both uplink and downlink channels in all the simulations.
}

\vspace{0.1in}
\subsection{Receiver Design}

{We aim to conduct the optimal detection based on the linear MIMO equation of (\ref{system_equ_uplk2}) using a SD}. For random channel fading coefficients, it is assumed that the rank of $\mathbf{H}$ is $K$ which is less than $J$. Hence, the rank-deficient linear equation in (\ref{system_equ_uplk2}) may not be solved by a standard SD \cite{FP1985}. In this work, we adopt the GSD proposed in \cite{Cui2005} by Cui and Tellamura in 2005. For self-containment, we sketch the derivation of the Cui-Tellamura GSD as follows.

Let $\lambda$ be a positive constant. Consider the Cholesky decomposition of the positive definitive matrix $\mathbf{Q}\triangleq \mathbf{H}^H \mathbf{H}+\lambda \textbf{I}_{J}$, i.e., $\mathbf{Q}=\mathbf{D}^H\mathbf{D}$, where $\mathbf{D}$ is an upper triangular matrix. Moreover, let $\mathbf{r}\triangleq(\mathbf{H}\mathbf{D}^{-1})^H \mathbf{y}$. For $M=4$, we have
\begin{equation}\label{Cui-GSD}
\begin{split}
\hat{\mathbf{b}}&=\arg \min_{\mathbf{b}\in \{1,-1\}^{2J}} \left( \|\mathbf{y}-\mathbf{H}\mathbf{u}\|^2+\lambda \mathbf{u}^H \mathbf{u}\right)\\
&=\arg \min_{\mathbf{b}\in \{1,-1\}^{2J}} \left ( \mathbf{y}^H\mathbf{y}-\mathbf{y}^H \mathbf{H} \mathbf{u} -\mathbf{u}^H\mathbf{H}^H\mathbf{y}+\mathbf{u}^H \mathbf{Q}\mathbf{u}\right )\\
&=\arg \min_{\mathbf{b}\in \{1,-1\}^{2J}} \|\mathbf{r}-\mathbf{D}\mathbf{u}\|^2.
\end{split}
\end{equation}
The derivation in (\ref{Cui-GSD}) shows that the above rank-deficient linear equation can be transformed to a full-rank one, which enables the use of a standard SD. Also, note that $\mathbf{u}=[\mu_1,\mu_2,\cdots,\mu_J]^T$ in (\ref{Cui-GSD}) is associated with $\mathbf{b}=[b_{1,1},b_{1,2},\cdots,b_{J,1},b_{J,2}]^T$ through $\mu_j=(b_{j,1}+ib_{j,2})/\sqrt{2}$ ($1\leq j \leq J$).

\vspace{0.2in}
\section{Comparisons of SCMA and DCMA}\label{section-IV}

In this section, we conduct numerical evaluations to compare DCMA and SCMA systems for $M=4$ in terms of their error rate performance and receiver complexity. {We are interested in comparing the optimal BERs of
both SCMA and DCMA in order to reveal the effect of their different diversity orders. We adopt the single tree-search (STS) based GSD \cite{Studer2008,Studer2010} for soft-input soft-output (SISO) decoding of DCMA system. A major advantage of STS-GSD is that it is capable of achieving BER approaching to that of maximum likelihood (ML) receiver with relatively low complexity. } For optimal detection of DCMA, we set $\lambda=1$ in GSD as suggested in \cite{Cui2005}.  In all simulations, we assume 1) Rayleigh fading channels, 2) perfect channel fading coefficients known to the receiver {(except for Subsection IV-B)}, and {3) uniform power allocation}. To study the error rate performances of CD-NOMA systems in coded block transmission, we define the system throughput as
\begin{equation}\label{throughput_equ}
\mathcal{T}\triangleq(J/K)R\log_2(M),
\end{equation}
where $R$ denotes the channel code rate. We consider two CD-NOMA system settings: 1) $K=4, J=6$ and  2) $K=5,J=10$. The indicator matrix in (\ref{ind_matrix_equ}) is used to construct the SCMA systems with the first CD-NOMA setting, whereas the indicator matrix below is used for SCMA systems with the second one.
\begin{equation}\label{ind_matrix_equ2}
\mathbf{F}=\left [
\begin{matrix}
1 & 1 & 1 &1 &0 &0 &0 &0 &0 &0\\
1 & 0 & 0 &0 &1 &1 &1 &0 &0 &0\\
0 & 1 & 0 &0 &1 &0 &0 &1 &1 &0\\
0 & 0 & 1 &0 &0 &1 &0 &1 &0 &1\\
0 & 0 & 0 &1 &0 &0 &1 &0 &1 &1
\end{matrix}
\right ].
\end{equation}	

\vspace{0.1in}
\subsection{Comparison of uncoded BER with perfect channel coefficients}

\begin{figure*}[htbp]
\centerline{
\subfloat[$(4\times6)$ CD-NOMA systems (downlink)]{\includegraphics[trim=2.2cm 8cm 2.2cm 8.5cm, clip=true, width=3in]{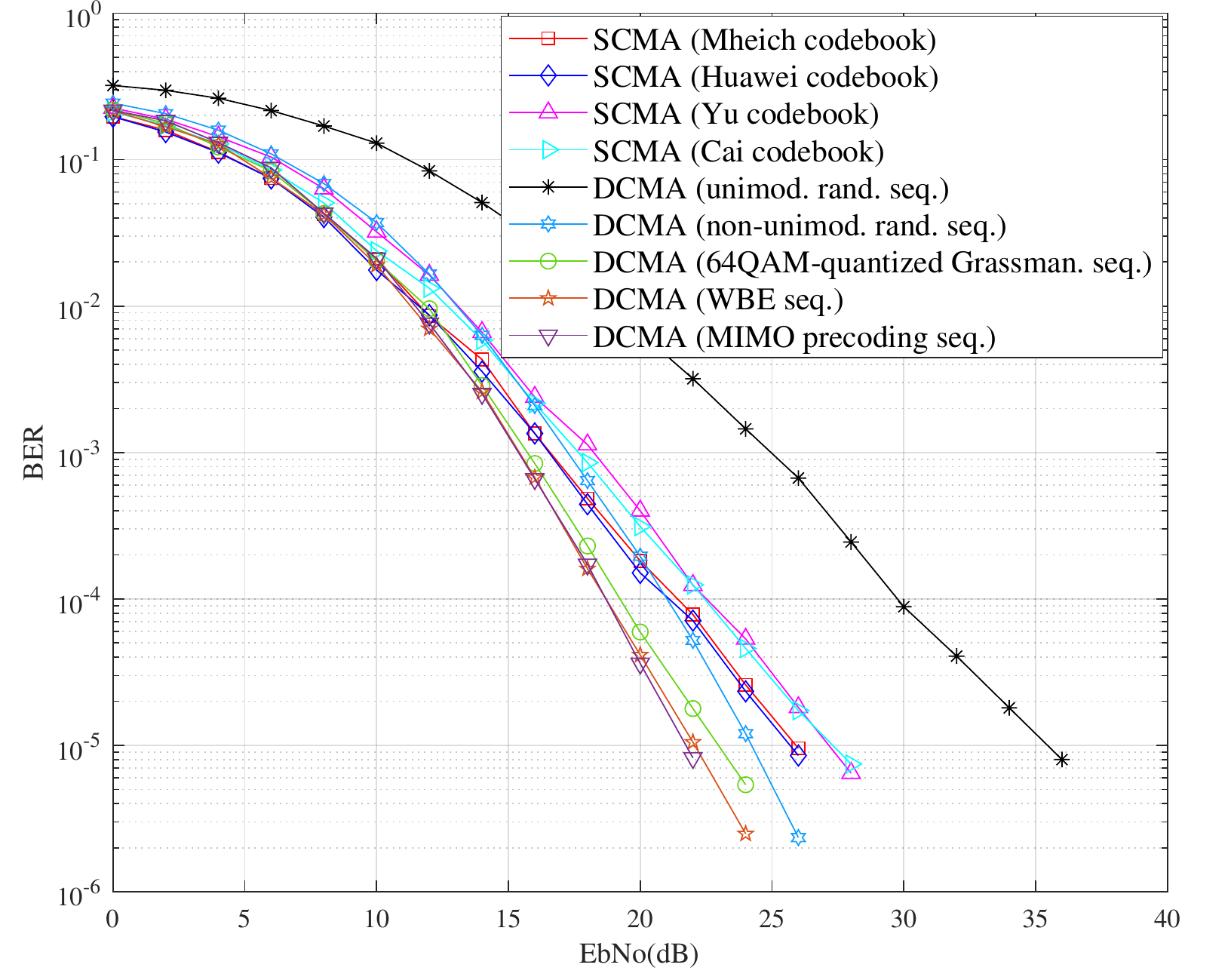}
\label{fig_ber_a}}
\hspace{-0.1in}
\subfloat[$(5\times10)$ CD-NOMA systems (uplink)]{\includegraphics[trim=2.2cm 8cm 2.2cm 8.5cm, clip=true, width=3in]{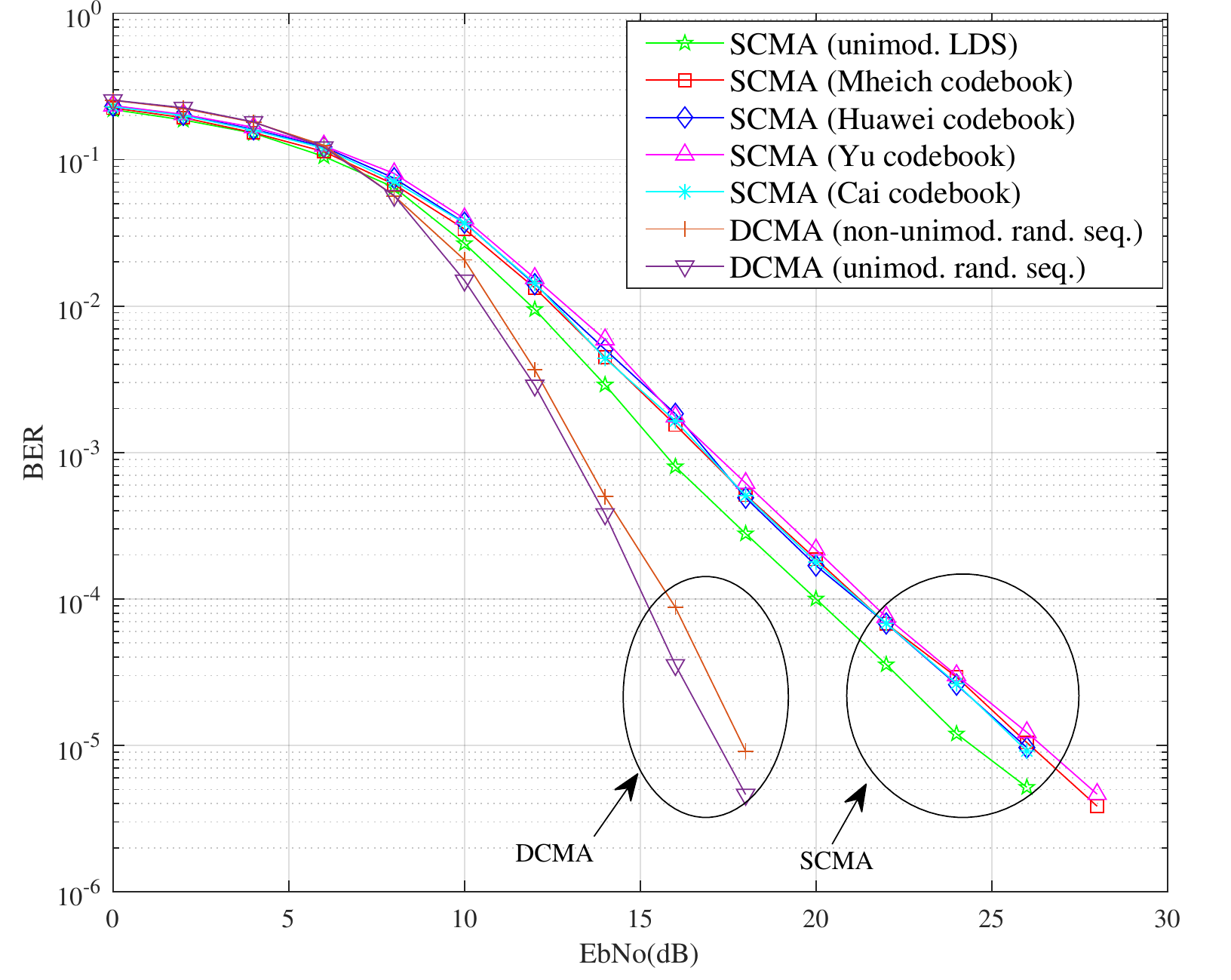}
\label{fig_ber_b}}
}
\caption{Uncoded BER comparison for CD-NOMA systems.}
\label{fig_BERComp}
\end{figure*}

Fig. \ref{fig_BERComp} compares the uncoded BER for DCMA and SCMA under different codebooks. For downlink transmission of $(4\times 6)$-DCMA, we adopt the following spreading matrix by \textit{Construction 3} (termed ``MIMO precoding seq." in Fig. \ref{fig_BERComp}-a):
\begin{equation}\label{DCMA4x6_codebook}
\left [
\begin{matrix}
1 & \theta_1 & \theta^2_1 &  \theta^3_1 &  \theta^4_1 &  \theta^5_1\\
1 & \theta_2 & \theta^2_2 &  \theta^3_2 &  \theta^4_2 &  \theta^5_2\\
1 & \theta_5 & \theta^2_5 &  \theta^3_5 &  \theta^4_5 &  \theta^5_5\\
1 & \theta_6 & \theta^2_6 &  \theta^3_6 &  \theta^4_6 &  \theta^5_6
\end{matrix}
\right ],
\end{equation}
where $L=7$ (as $J=6$) and hence $\theta_1=\exp(i2\pi/7),\theta_2=\exp(i4\pi/7),\theta_5=\exp(i10\pi/7),\theta_6=\exp(i12\pi/7)$. That is, we select rows 1, 2, 5, 6 from $\Theta$. Although we tried some other selection schemes of rows from $\Theta$, no major improvement has been observed in terms of the BER performance. We also adopt a WBE spreading matrix (termed ``WBE seq." in Fig. \ref{fig_BERComp}-a) generated by the iterative construction method in \cite{Ulukus2001}. The ``64QAM-quantized Grassman. seq." are obtained from \cite{NOMA2018}. In addition, we simulate unimodular\footnote{Dense sequences with random phases and identical magnitude.} and non-unimodular\footnote{Random dense sequences subject to Gaussian normal distribution.} random dense sequences (termed ``unimod. rand. seq." and ``non-unimod. rand. seq.", normalised with identical sequence energy for all the users). For SCMA, we consider the Mheich codebook in \cite{zeina2019gam}, the Huawei codebook in \cite{nikopour2013scma}, the Yu codebook in \cite{yu2018star}, the Cai codebook in \cite{cai2016multi}. One can see that 1) DCMA generally leads to significantly improved BER (with steeper BER curves) compared to SCMA due to its capability of achieving full DO; 2) The best BER performance is attained by DCMA with MIMO precoding sequences which enjoys about 4 dB gain at BER of $10^{-5}$; 3) The only exception is DCMA with unimodular random sequences under which the superposed signals from multiple users may be more likely to be nulled to a very small number close to zero; 4) The four SCMA codebooks display similar BER slopes which are less steeper than that of DCMA as SCMA suffers from DO less than $K$.

For uplink $(5\times 10)$-DCMA, as all the channel fading coefficients associated to each user are random and independent (i.e., Rayleigh fading channel), the structure of a carefully designed codebook may not be held after passing through the wireless channels. For this reason, we only consider DCMA with  {``unimod. rand. seq." and ``non-unimod. rand. seq."} We also consider ``unimodular LDS" which refers to LDS codebooks whose nonzero elements take identical magnitude. Similarly to the downlink $(4\times 6)$-NOMA case, Fig. \ref{fig_BERComp}-b shows that 1) DCMA with {``unimod. rand. seq."} benefits from full DO and outperforms the five SCMA codebooks for at least 8 dB at BER below $10^{-5}$; 2) As pointed out in \textit{Theorem \ref{main-theorem}}, DCMA with {``unimod. rand. seq." (compared to that with ``non-unimod. rand. seq.")} in uplink channels enjoys the best BER performance; 3) As far as SCMA is concerned, ``unimodular LDS" outperforms the other four SCMA codebooks as the former gives rise to minimum rate of single-error patterns (as stated in \textit{Remark \ref{rmk4SCMA}}) which are the dominant error source.

{
\subsection{Comparison of uncoded BER with channel estimation error (CEE)}

\begin{figure*}[htbp]
\centerline{
\subfloat[{$(4\times6)$ CD-NOMA systems}]{\includegraphics[trim=3.25cm 9.25cm 3.25cm 9.25cm, clip=true, width=3.2in]{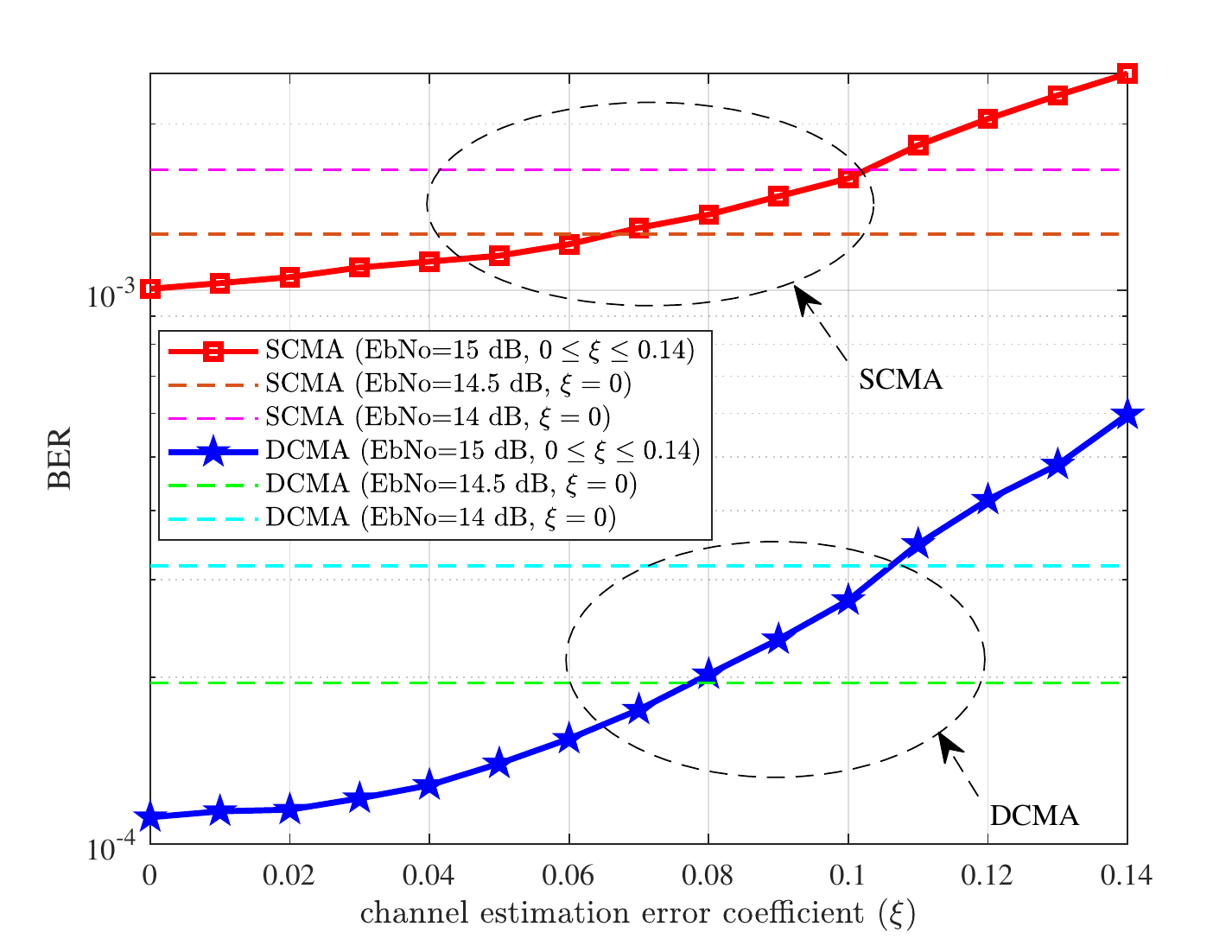}
\label{fig_ber_a2}}
\hspace{-0.1in}
\subfloat[{$(5\times10)$ CD-NOMA systems}]{\includegraphics[trim=3.25cm 9.25cm 3.25cm 9.25cm, clip=true, width=3.2in]{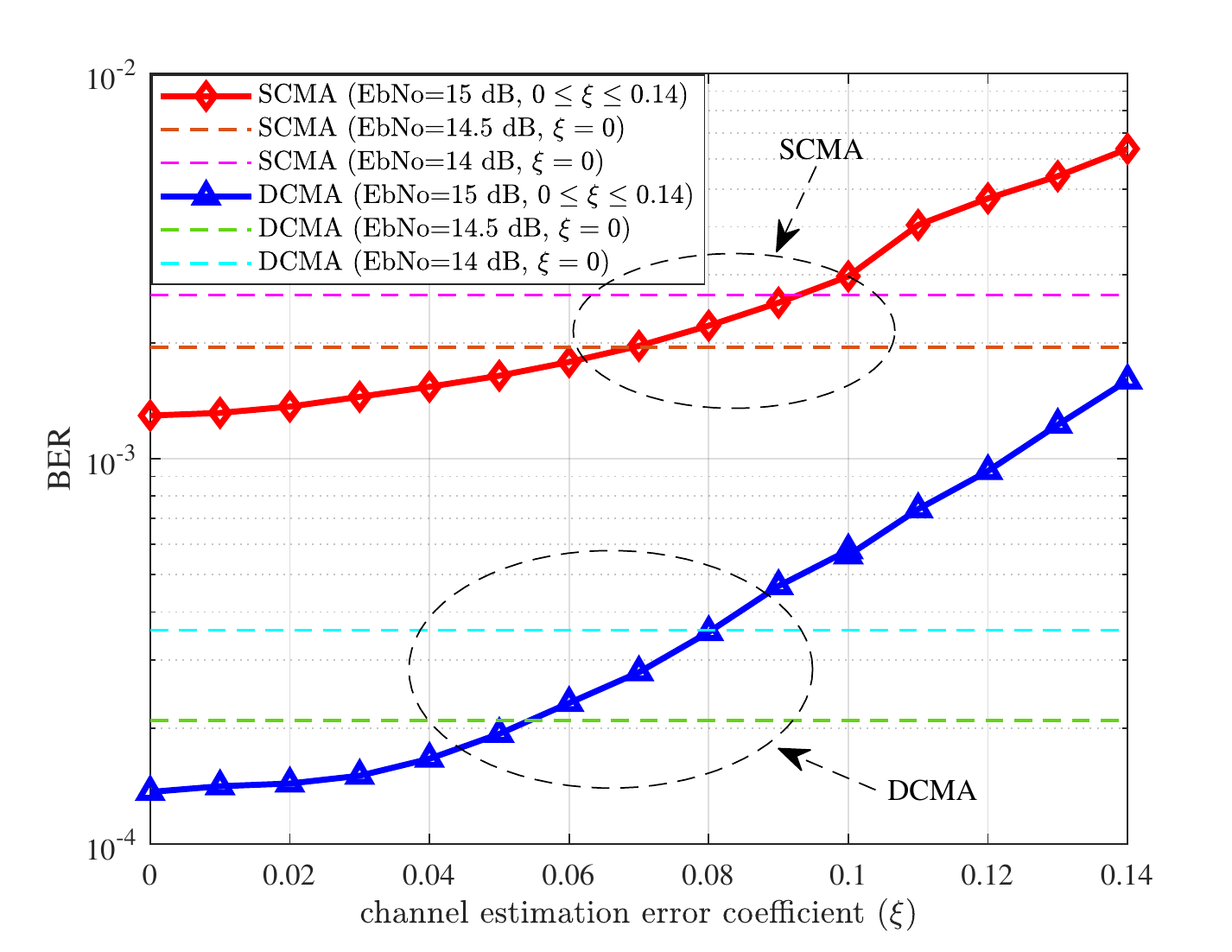}
\label{fig_ber_b2}}
}
\caption{{Uncoded BER comparison with CEE coefficient $\xi$ in uplink channels for EbNo at 15 dB.}}
\label{fig_BERComp_chanest}
\end{figure*}

In practical wireless communication systems, it may be difficult to obtain perfect channel coefficients. Hence, it is enlightening to compare the BERs of CD-NOMA systems with channel estimation errors (CEEs). Let us consider the channel fading vector $\mathbf{h}_j$ of user $j$ ($1\leq j \leq J$). Formally, the estimated channel fading vector of user $j$ is modelled as
\begin{equation}
\hat{\mathbf{h}}_j=\mathbf{h}_j\cdot (1+\xi \cdot \triangle_j),
\end{equation}
where $0<\xi\ll 1$ is called the (normalized) CEE coefficient and $\triangle_j$ is a complex-valued random variable which is uniformly distributed over the unitary circle $|x|\leq 1$.

In Fig. \ref{fig_BERComp_chanest}, we compare the uncoded BERs of SCMA and DCMA with $\xi\in [0,0.14]$ in uplink channels with EbNo of 15 dB, which are denoted by ``BER(SCMA, EbNo=15 dB, $0\leq \xi \leq 0.14$)" and ``BER(DCMA, EbNo=15 dB, $0\leq \xi \leq 0.14$)", respectively. It is noted that a nonzero CEE may lead to deterioration of BER which is similar to the effect of EbNo decrease. Thus, we have also simulated the BERs at EbNo of 14 dB and 14.5 dB but with $\xi=0$. By comparing $\xi_1,\xi_2$  which satisfy
\begin{equation}\label{BER_chanesterr}
\begin{split}
\text{BER(SCMA, EbNo=15 dB}, \xi_1)& =\text{BER(SCMA, EbNo=14.5 dB}, \xi=0),\\
\text{BER(DCMA, EbNo=15 dB}, \xi_2)& =\text{BER(DCMA, EbNo=14.5 dB}, \xi=0),
\end{split}
\end{equation}
respectively, one can decide which system is more resilient to CEE.  For the BERs of the $(4\times 6)$ CD-NOMA systems shown in Fig. \ref{fig_BERComp_chanest}-a, we have $\xi_1\approx 0.065,\xi_2\approx 0.08$, indicating that DCMA is more resilient to CEE. The same assertion holds by plugging ``EbNo=14 dB" into the right-hand-side of (\ref{BER_chanesterr}). For the BERs of the $(5\times 10)$ CD-NOMA systems shown in Fig. \ref{fig_BERComp_chanest}-b, however, SCMA is more resilient as $\xi_1\approx0.07,\xi_2\approx0.055$. Despite the above observations, no CD-NOMA is drastically sensitive to CEE.
}

\vspace{0.1in}
\subsection{Comparison of BLER}

 \begin{figure*}[htbp]
  \centering
  \includegraphics[width=6in]{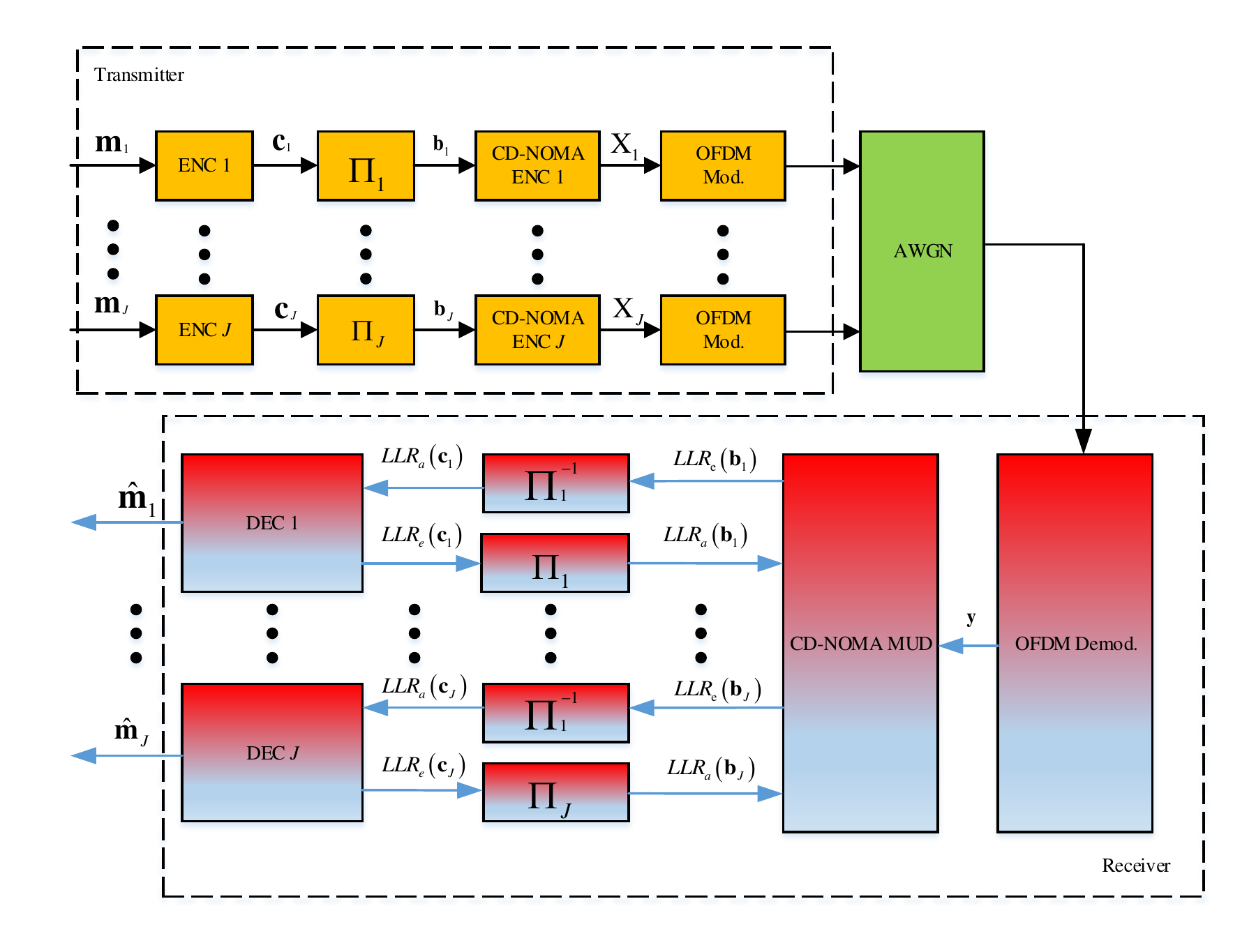}\\
  \caption{A coded CD-NOMA system model in uplink channel with $J$ users with iterative Turbo receiver.}
  \label{system_model2}
\end{figure*}

\begin{figure*}[htbp]
\centerline{
\subfloat[LDPC coded $(4\times6)$-NOMA systems]{\includegraphics[trim=2.25cm 8cm 2.25cm 8.45cm, clip=true, width=3.2in]{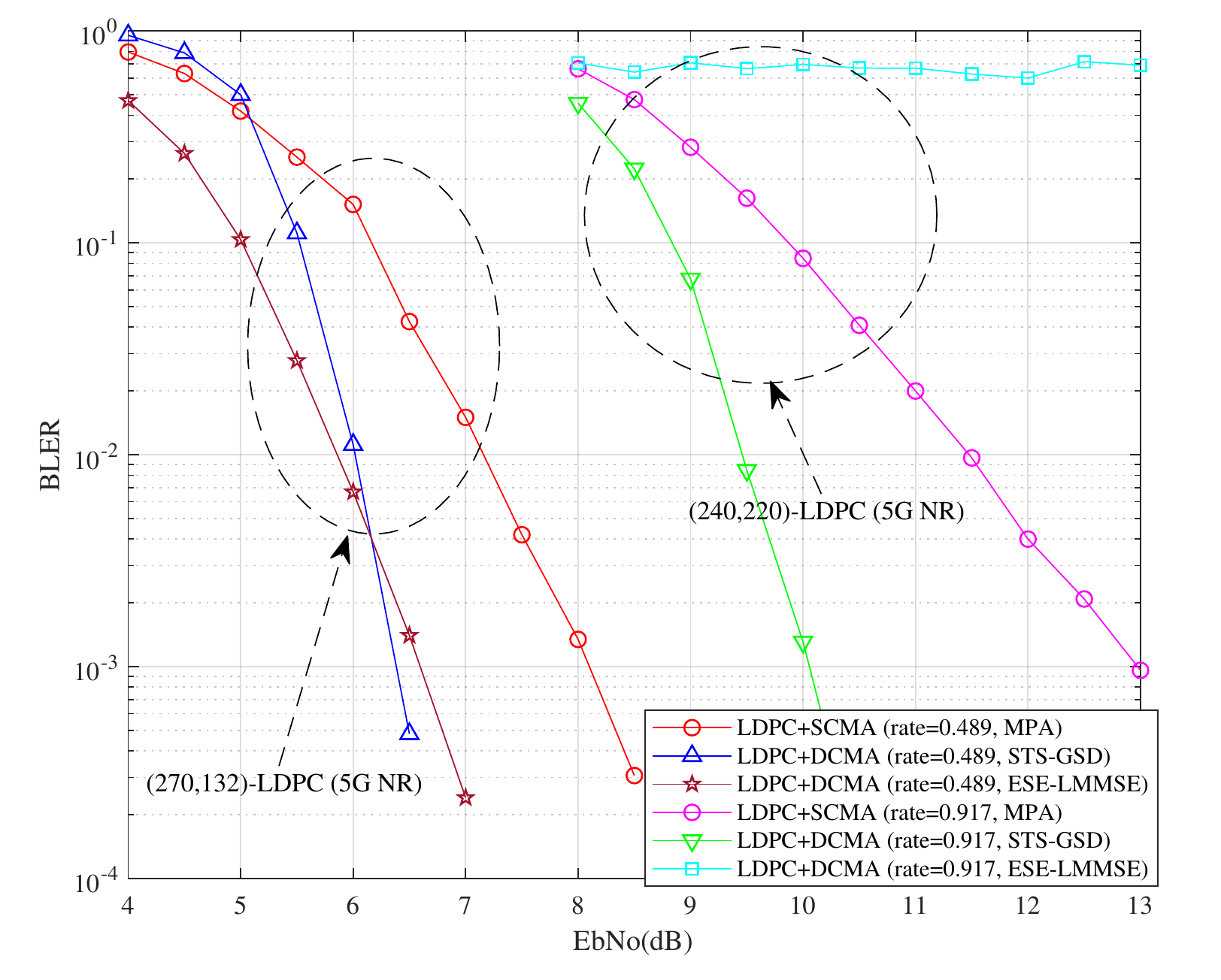}
\label{fig_bler_a}}
\hspace{-0.1in}
\subfloat[LDPC coded $(5\times10)$-NOMA systems]{\includegraphics[trim=2.25cm 8cm 2.25cm 8.45cm, clip=true, width=3.2in]{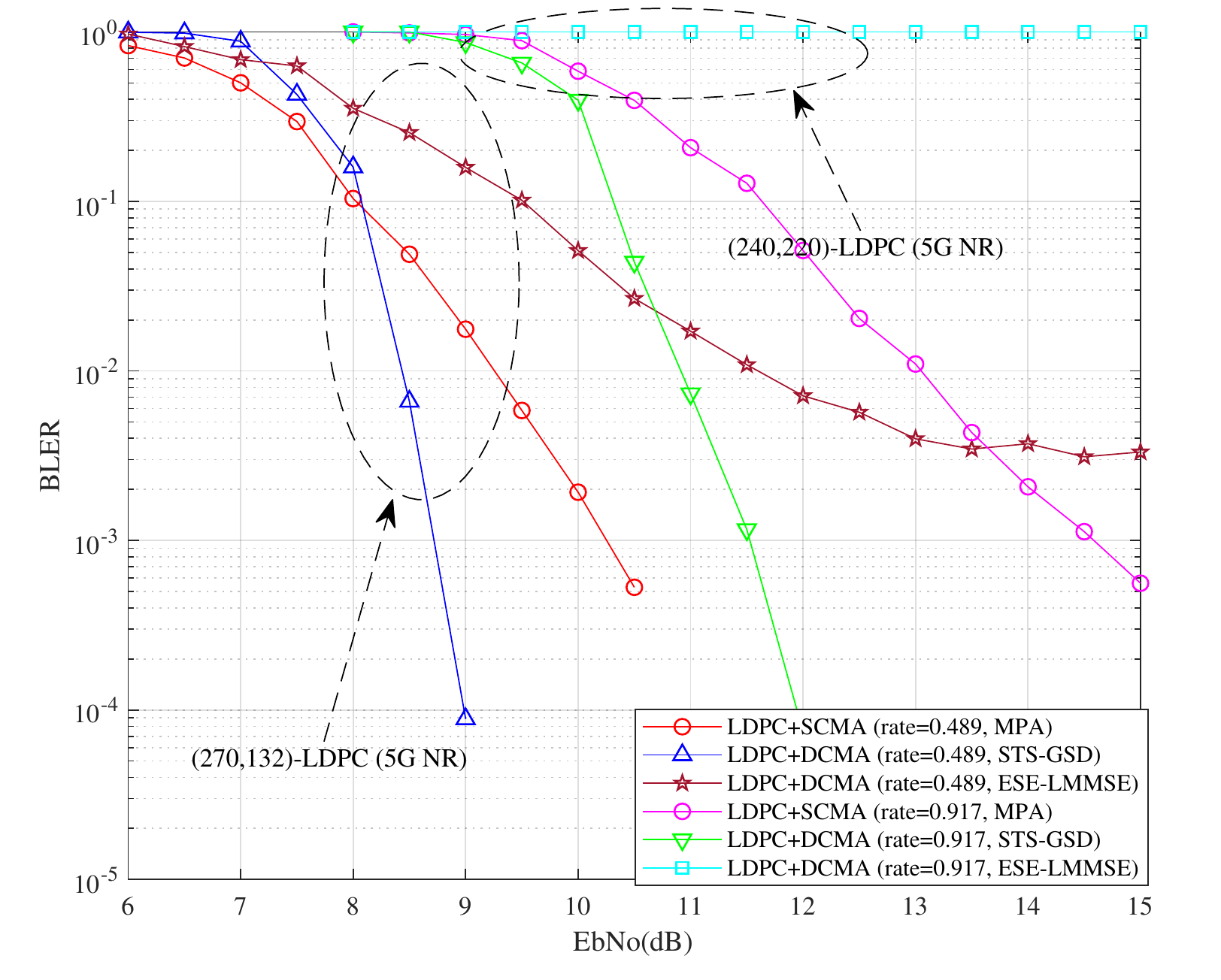}
\label{fig_bler_b}}
}
\caption{BLER comparison for LDPC coded CD-NOMA systems under iterative Turbo receiver in uplink Rayleigh fading channels. The outer-loop Turbo iterations are set to be 3 for both DCMA with STS-GSD detector and SCMA with MPA detector and 20 for DCMA with ESE-LMMSE detector.}
\label{fig_BLERComp}
\end{figure*}

In this subsection, we compare the block error rate (BLER) performance of different CD-NOMA systems.
Fig. \ref{system_model2} presents a coded CD-NOMA system structure in uplink channel. At the transmitter side, the information bits of  user $j$ are first encoded by a channel encoder (denoted by ``ENC"), followed by a random interleaver (denoted by $\Pi_j$). Then the transmitter carries out CD-NOMA encoding as well as OFDM modulation. At the receiver side, after OFDM demodulation, Turbo decoding is carried out between CD-NOMA MUD and channel decoder (denoted by ``DEC") by iteratively exchanging soft information in the form of log-likelihood ratio (LLR) including \textit{a priori} $LLR_a$ and \textit{a posteriori} $LLR_e$ (extrinsic).

Fig. \ref{fig_BLERComp} compares the BLER performance of the LDPC coded CD-NOMA systems. For each CD-NOMA setting and considering the short-packet nature of MTC networks, we apply two 5G NR LDPC codes, as specified in TS38.212 \cite{ts38212}, with rates of $132/270$ and $220/240$, respectively. For example, when the first LDPC code is used, each block consists of 132 bits and 270 bits before and after encoding, respectively. 
To examine the BLER performance of low-complexity receiver for DCMA, we also consider LMMSE detector \cite{Wang1999,Tuchler2002,Li2009,Guo2011,Haselmayr2017} associated to the so-called elementary signal estimator (ESE) \cite{Li2006}. Such an ESE-LMMSE detector can be efficiently implemented based on vector/scalar Gaussian approximation \cite{Haselmayr2017}. The key idea of  the ESE-LMMSE detector is to recursively update the means and covariance matrices of CD-NOMA symbols by leveraging the \textit{a priori} LLR inputs from the channel decoders. The outer-loop iterations are set to 3 for both DCMA with STS-GSD detector and SCMA with MPA detector and 20 for DCMA with ESE-LMMSE detector. As uplink channel is considered, we adopt unimodular dense sequences for DCMA and ``Unimodular LDS" codebooks for SCMA for the best error rate performances. We have the following key observations:
\begin{enumerate}
\item For the two NOMA settings, the DCMA systems under STS-GSD detector with rate $R=220/240$ achieve about 3 dB gain over the corresponding SCMA counterparts at BLER of $10^{-3}$. In this case, as the rate is high, little coding gain can be harvested and hence the BLER gain is mainly due to the full DO of DCMA. When the lower rate of $R=132/270$ (i.e,. steeper BLER curves due to higher coding gain), still 1.5 dB gain can be attained by DCMA.
\item For the $(4\times6)$-DCMA system, it is interesting to see ESE-LMMSE detector works well when $R=132/270$, in which $\mathcal{T}=1.467$ bits. In particular, in this case, ESE-LMMSE detector enjoys lower BLER (compared to that of STS-GSD detector) for $\text{Eb/No}$ no greater than 6 dB. However, its BLER performance starts to deteriorate for $(5\times10)$-DCMA system with $R=132/270$ (i.e., $\mathcal{T}=1.956$ bits). As a matter of fact,  the ESE-LMMSE detector seems incompetent in exploiting the full DO property of DCMA\footnote{The situation may be improved for sufficiently long channel code, but the investigation is beyond our research focus of this paper.} as its BLER curve of ESE-LMMSE detector is worse than that of SCMA and exhibits some error floor in high SNR region (see Fig. \ref{fig_BLERComp}-b). At $R=220/240$, we have $\mathcal{T}=2.750$ bits and $\mathcal{T}=3.667$ bits for the two different CD-NOMA settings, under which however ESE-LMMSE detectors for DCMA fail to work. In comparison, for all the throughputs considered in Fig. \ref{fig_BLERComp}, the BLER curves can converge well for STS-GSD detector based DCMA and MPA detector based SCMA.
\end{enumerate}

\vspace{0.1in}
\subsection{Comparison of complexity}

In this subsection, we compare the complexities of the STS-GSD detector\footnote{A major advantage of the ESE-LMMSE detector is its low implementation complexity. By assuming $J$ and $K$ are on the same order, the complexity of the ESE-LMMSE detector is estimated to be $O(J^2)$ \cite{Li2009,Guo2011}. That being said, as shown in the preceding subsection, the ESE-LMMSE detector may not work for a high system throughput which is however an essential requirement for MTC networks.} for DCMA (including the pre-SD matrix calculations) and the MPA detector for SCMA using floating-point (FLOP) operations, each of which refers to either a complex multiplication or a complex addition. In fact, the MPA detector for SCMA has complexity of $O(N_{\text{iter}}KM^{d_c}d_c^2)$ \cite{Wei2017}, where $N_{\text{iter}}$ refers to the number of MPA iterations. In the simulations, for decoding convergence, we set $N_{\text{iter}}$ to be 5 and 10 for $(4\times 6)$-SCMA and $(5\times 10)$-SCMA, respectively. {Here, $N_{\text{iter}}$ is selected as the minimum integer at which the decoding of MPA attains convergence.} For standard SD, the expected complexity is in proportional to the average number of visited nodes of each level during the tree search, which is roughly cubic in the number of binary variables to be solved \cite{Hassibi2005}.  ``The average number of visited nodes" may increase in low SNR region or large-scale DCMA systems. To proceed, we define normalized complexity as follows:
\begin{equation}
\text{Normalized Complexity}\triangleq \frac{\text{Number of FLOPs}}{(J\log_2 M)^3}.
\end{equation}

\begin{figure*}[ht]
  \centering
  \includegraphics[trim=3.5cm 9.2cm 3.75cm 10cm, clip=true, width=4.5in]{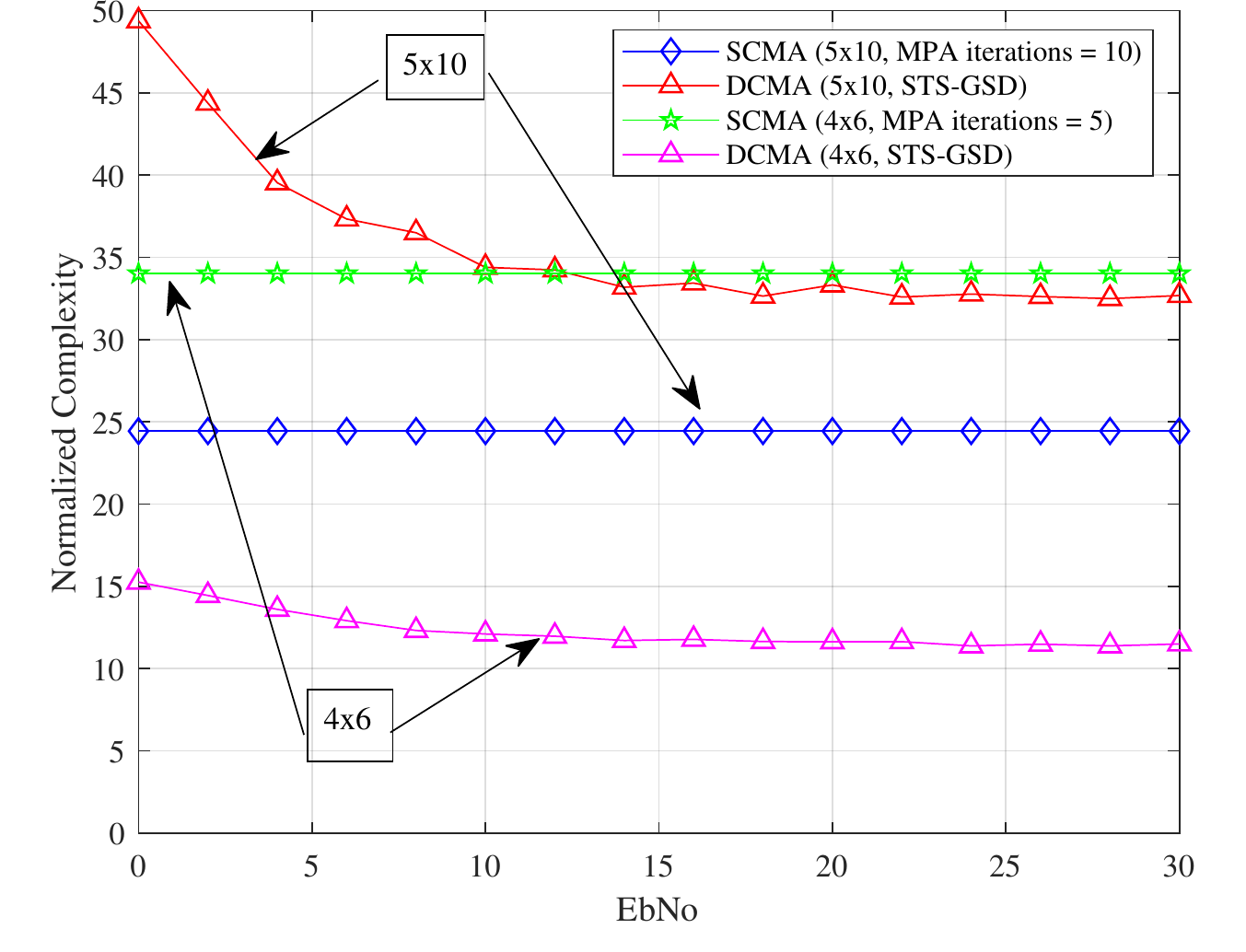}
  \caption{Complexity comparison for CD-NOMA systems ($M=4$).}
  \label{Fig_FLOPComp}
\end{figure*}

For the $(4\times 6)$ CD-NOMA setting, Fig. \ref{Fig_FLOPComp} indicates that the normalized complexity of DCMA detector is only about $35\%$ of that of SCMA detector. The $(5\times 10)$-DCMA system, on the other hand, has about $33\%$ additional complexity than that of the SCMA counterpart due to increased average number of visited nodes in the tree search. When uplink transmission is concerned, as the decoding is conducted at a {base station} receiver, the complexity increase for the $(5\times 10)$-DCMA system may be durable.

\vspace{0.2in}
\section{Conclusions and Future Works}\label{section-V}
CD-NOMA is an emerging paradigm to support efficient information exchange over massive number of machine-type communication devices. In this paper, we have carried out a comparative study for the two \textit{overloaded} CD-NOMA schemes, i.e., SCMA and DCMA. SCMA is named due to its sparse codebooks which allow the use of MPA detector, whereas DCMA bears some similarity to legacy CDMA as dense codebooks/sequences are adopted. We have considered CD-NOMA transmitted over an OFDM system, where every subcarrier receives an independent Rayleigh fading gain.

Despite numerous research attempts on SCMA in recent years, our analysis in Section III for the PEP has shown that SCMA suffers from relatively small DO, which is a bottleneck for significant performance enhancement. By contrast, the error rate performance of DCMA outperforms that of SCMA as the former enjoys full DO by spreading every user's data symbols over all the subcarriers. Over uplink Rayleigh fading channels, we have proved and validated through numerical simulations (in Section IV) that unimodular sequences appear to be the optimal codebooks for DCMA with $M=4$ as they lead to the largest minimum product distance in Rayleigh fading channels. We have also found that unimodular LDS (i.e., sparse sequences whose nonzero elements possess identical magnitude) lead to optimal SCMA codebooks in terms of their single-error PEPs.  For downlink Rayleigh fading channels, we have suggested to use a number of transform matrices from the areas of lattice constellation shaping and MIMO precoding for good dense sequences.

We have found that the selection of a proper detector plays a key role in exploiting the full DO property of DCMA. In this paper, we have adopted a non-linear DCMA detector based on GSD (with STS-GSD for SISO detection), whose superior BER and BLER performance have been demonstrated in Section IV. Although the ESE-LMMSE detector has an advantage of relatively low complexity, it is interesting to reveal that the BLER of the corresponding DCMA may not converge when the system throughput\footnote{Definition of system throughput can be found in (\ref{throughput_equ}).} is larger than 2. By counting the FLOPs operations at the receivers for both the $(4\times 6)$ and $(5\times 10)$ CD-NOMA systems, we have shown that the STS-GSD detector for DCMA has a complexity comparable to that of the MPA detector for SCMA. 

\textit{Future Directions}: 1) During this research, we also tried to use a larger NOMA setting than the ones considered in the current work. However, we observed that the complexity of the STS-GSD detector increases rapidly when a larger NOMA setting is adopted. In this case, SCMA may be more attractive in terms of receiver complexity. It is therefore interesting to develop a low-complexity DCMA SISO detection algorithm whose error rate performance is comparable to that of the STS-GSD detector. Some advanced MIMO detectors \cite{Goldberger2011,Cespedes2014} may be a good starting point for breakthrough. 2) To improve the performance of SCMA, besides adopting a strong channel code, it is worthy to investigate spatial coupling aided SCMA (SC-SCMA), where SC is an effective approach for improved BP decoding threshold in coding theory \cite{Takeuchi2015,Felstrom1999}. Further research is needed in understanding the performances of SC-SCMA in different channel conditions.

\bibliographystyle{IEEEtran}
\bibliography{bib}

\end{document}